\documentclass[11pt]{article}
\pdfoutput=1
\usepackage{graphicx}
\usepackage{hyperref}
\usepackage{epsfig}
\usepackage{amsmath}
\usepackage{amsthm}
\usepackage{amssymb}
\usepackage{multirow}
\usepackage{color}
\usepackage[nosort]{cite}
\usepackage{hyperref}
\usepackage[fontsize=11.35pt]{scrextend}
\usepackage{braket,mathrsfs,slashed}
\usepackage{float}
\usepackage{setspace}
\usepackage[caption=false]{subfig}
\usepackage[export]{adjustbox}
\usepackage{enumitem}

\newlength{\abstractwidth}
\newcommand{\be}{\begin{equation}}
\newcommand{\ee}{\end{equation}}

\renewcommand{\title}[1]{\vbox{\center\bf{\Large{#1}}}\vspace{5mm}}
\renewcommand{\author}[1]{\vbox{\center#1}\vspace{5mm}}
\newcommand{\address}[1]{\vbox{\center\em#1}}
\newcommand{\email}[1]{\vbox{\center\tt#1}\vspace{5mm}}

\usepackage{dsfont}

\usepackage{rotating}
\usepackage{tikz-feynman}

\usepackage{dsfont}
\usepackage[paper=a4paper,margin=1.5cm]{geometry}
\usepackage[utf8]{inputenc}
\usepackage{rotating}
\usepackage{soul}

\usepackage{mathrsfs} 
\usepackage{bbm}
\usepackage{etoolbox} 
\usepackage{multirow}

\allowdisplaybreaks

\hypersetup{pdfstartview=FitV,colorlinks=true,linkcolor=midblue,citecolor=midblue,filecolor=midblue,urlcolor=midblue}

\definecolor{midblue}{rgb}{0,0,0.5}

\begin{document}
	
\newgeometry{top=3.1cm,bottom=3.1cm,right=2.4cm,left=2.4cm}
	
\begin{titlepage}
\begin{center}
\hfill \\
\vskip 0.5cm

\title{Strict renormalizability\\[1.5mm] as a paradigm for fundamental physics}

\author{\large Luca Buoninfante}
			
\address{High Energy Physics Department, Institute for Mathematics, Astrophysics,\\
and Particle Physics, Radboud University, Nijmegen, The Netherlands}
\email{\rm \href{mailto:luca.buoninfante@ru.nl}{luca.buoninfante@ru.nl}}

\end{center}

\begin{abstract}
An important theoretical achievement of the last century was the realization that strict renormalizability can be a powerful criterion to select Lagrangians in the framework of perturbative quantum field theory. The Standard Model Lagrangian (without gravity) is strictly renormalizable from a perturbative point of view. On the other hand, the inclusion of gravity seems not to respect this criterion, since general relativity is perturbatively non-renormalizable. The aim of this work is to provide concrete evidence that strict renormalizability is still a valid criterion even when applied to gravity. First, we show that adding quadratic curvature terms to the Einstein-Hilbert action gives rise to a strictly renormalizable theory known as quadratic gravity. Second, we argue that this unique theory represents the most conservative approach to quantum gravity and, at the same time, is highly predictive, as it can explain new physics beyond general relativity already in the sub-Planckian regime. In particular, it provides one of the best fits to the CMB anisotropies via Starobinsky inflation and makes sharp cosmological predictions that can be tested in the near future. Finally, we comment on the (super-)Planckian regime and conclude with a historical note.
\end{abstract}

\end{titlepage}
	
{	
\hypersetup{linkcolor=black}	
\tableofcontents
}

\baselineskip=17.63pt

	

\newpage
	
\section{The QFT framework and its guiding principles}

The framework of relativistic quantum field theory (QFT) has provided useful tools for describing electromagnetic, weak, and strong interactions at the fundamental level. One of its major successes is that with very few assumptions it severely constrains the Lagrangians, thus making the Standard Model (SM) of particle physics very predictive~\cite{Iliopoulos:2025fhr}.

The main ``guiding principles'' that are usually required are the following:
\begin{itemize}

\item \textbf{Locality:} The bare Lagrangian depends only polynomially on the derivatives, i.e.
\begin{equation}
\mathcal{L}\equiv \mathcal{L}\left(\varphi,\partial\varphi,\partial^2\varphi,\dots,\partial^n\varphi\right)\,,\qquad n<\infty\,,
\end{equation}
where $\varphi$ is some tensorial or spinor field.

\item \textbf{Symmetries:} Actions can be invariant under  spacetime and internal (global and gauge) symmetry groups.

\item \textbf{Unitarity:} Quantum probabilities are conserved, which mathematically means that the evolution operator is unitary, for example in terms of the $S$-matrix we have $S^\dagger S=\mathds{1}.$

\item \textbf{Strict renormalizability:} The ultraviolet (UV) behavior of the theory is governed by dimensionsless interaction couplings.

\end{itemize}

Here by ``renormalizability'' we mean ``perturbative renormalizability'', i.e. we refer to Dyson's criterion according to which QFTs are defined to be renormalizabile when the interaction couplings have non-negative dimensions in units of mass, with respect to the power counting that controls the UV behavior of the theory~\cite{Dyson:1949bp,Dyson:1949ha,Sakata:1952rq}. In this case, the number of counterterms and physical parameters needed to absorb the UV divergences in perturbation theory is finite. We can distinguish two subclasses:\footnote{See also section 2.A.4 in Ref.~\cite{Basile:2024oms} for a pedagogical discussion.} strictly renormalizable QFTs in which the couplings that control the UV behavior are dimensionless; super-renormalizable QFTs in which these couplings have positive mass dimension. Furthermore, (perturbatively) non-renormalizable QFTs are defined such that the UV behavior is governed by at least one coupling that has negative mass dimension. In the latter case, UV divergences can still be renormalized~\cite{Anselmi:1994ry,Gomis:1995jp} but an infinite number of counterterms is needed, therefore predictivity may be lost at very high energies because a finite number of experiments will never be able to measure an infinite number of independent parameters.

Once we specify the number of spacetime dimensions and the types of fields and symmetries, all together the guiding principles listed above turn out to be very restrictive. Locality excludes all possible non-polynomial (i.e. quasi-local and non-local) differential operators as terms in a bare Lagrangian; in particular, the SM bare Lagrangian contains first- and second-order derivatives. Poincaré/Lorentz symmetry, among other things, tells us that in a covariant Lagrangian the number of time and space derivatives acting on a field must be the same. Gauge symmetries, despite being redundancies in the number of degrees of freedom,  are very important to constrain the form of the interactions. Additionally, the requirement of gauge anomaly cancellation restricts the number and type of fields. Unitarity does not directly constrain the functional form of the Lagrangian but tells us about the signs of the probabilities and the arrow(s) of causality. For example, in the SM the unitarity condition (i.e. the optical theorem) is compatible with physical states having positive (squared) norms and propagators prescribed with the causal Feynman shift. Finally, strict renormalizability excludes the possibility that dimensionful couplings control the UV behavior of a theory. The SM Lagrangian, in which we also include mass terms for neutrinos via Yukawa interactions with the Higgs, falls into the class of strictly renormalizable QFTs.

Note that if we do not require this last criterion, we would have an infinite number of possible terms in a Lagrangian, which would significantly decrease the predictive power of the corresponding theory. Let us give some examples to show how strict~renormalizability works.

\medskip

\textbf{Example 1: Electromagnetic interaction.} Imagine that we want to write a Lagrangian that describes the dynamics of photons, electrons and positrons, including self- and mutual interactions. If we require locality, Poincaré symmetry, $U(1)$ gauge symmetry and unitarity, \textit{but} not strict renormalizability, we would write\footnote{In this work we always work in four spacetime dimensions ($D=1+3$), adopt the mostly positive convention for the metric signature ($-+++$), and choose the Natural units system ($\hbar=1=c$).}
\begin{equation}
\begin{aligned}
\mathcal{L}_{\rm em}=&-\frac{1}{4}F_{\mu\nu}F^{\mu\nu}-\bar{\psi}\left(i\gamma^\mu\partial_\mu +m\right)\psi-e\bar{\psi}\gamma^\mu A_{\mu}\psi\\
&+f \bar{\psi}\left[\gamma^\mu,\gamma^\nu\right] F_{\mu\nu}\psi+\lambda \left(F_{\mu\nu}F^{\mu\nu}\right)^2+\cdots\,,
\end{aligned}
\end{equation}
where $F_{\mu\nu}=\partial_{\mu}A_\nu -\partial_\nu A_\mu$ and the dots represent all infinite terms that are compatible with the stated requirements. The mass dimensions of the interaction couplings are
\begin{equation}
[e]=0\,,\qquad [f]=-1\,,\qquad [\lambda]=-4\,,\,\dots\,,
\end{equation}
and the other infinitely many couplings have negative mass dimensions.

If we also impose strict renormalizability as an additional requirement, we would only write
\begin{equation}
\mathcal{L}_{\rm QED}=-\frac{1}{4}F_{\mu\nu}F^{\mu\nu}-\bar{\psi}\left(i\gamma^\mu\partial_\mu +m\right)\psi-e\bar{\psi}\gamma^\mu A_{\mu}\psi\,,
\end{equation}
and nothing more. This is the Lagrangian of quantum electrodynamics (QED) which corresponds to a strictly renormalizable QFT~\cite{Tomonaga:1946zz,Tomonaga:1948zz,Schwinger:1948yk,Schwinger:1948yj,Schwinger:1948iu,Feynman:1949zx,Feynman:1950ir,Dyson:1949bp,Dyson:1949ha}.

\medskip

\textbf{Example 2: Electroweak interaction.} The strict renormalizability criterion can also help us to understand what to do when we would like to find a new renormalizable Lagrangian that reproduces a non-renormalizable one in some low-energy regime. A famous example is Fermi's theory of beta decay~\cite{Fermi:1933jpa} which is perturbatively non-renormalizable, since the lowest order interaction term is given by $G_{\rm F} (\bar{\psi}\psi)^2,$ where $G_{\rm F}\simeq (245 \,{\rm GeV})^{-2}$ and $[G_{\rm F}]=-2.$ Indeed, this theory breaks down at energies of the order of $\mathcal{O}(100)\,\text{GeV}$. However, it is well known that
the addition of the gauge bosons $W^{\pm}$, $Z^0$ and the Higgs boson $H$, supplemented by the mechanism of spontaneous symmetry breaking~\cite{Englert:1964et,Higgs:1964pj,Guralnik:1964eu}, gives rise to the Glashow-Weinberg-Salam electroweak theory~\cite{Glashow:1961tr,Weinberg:1967tq,Salam:1968rm}, which is strictly renormalizable~\cite{tHooft:1971akt,tHooft:1971qjg}.

\medskip

\textbf{Example 3: Strong interaction.} In the case of the strong force, the strict renormalizability paradigm does not actually suggest adding new particles, but rather helps us to identify the correct degrees of freedom. In fact, the low-energy dynamics can be effectively described in terms of pions whose Lagrangian is perturbatively non-renormalizable. However, at a more fundamental level the strong interaction is successfully described by quantum chromodynamics (QCD)~\cite{Fritzsch:1972jv,Fritzsch:1973pi}, which is a strictly renormalizable QFT according to which the fundamental degrees of freedom are gluons and quarks whose interactions are asymptotically~free~\cite{Politzer:1973fx,Gross:1973id}.

\medskip

\textbf{Remark.} It is important to clarify that strict renormalizability as a selection principle for fundamental Lagrangians is not limited to perturbative QFT and (non-convergent) asymptotic expansions. It also tells us which is the Lagrangian we need to plug into a path integral to perform fully non-perturbative analyses, e.g. study of soliton-like solutions, tunneling probabilities, lattice simulations, etc. A priori this does not imply that a strictly renormalizable local Lagrangian is always suitable to describe physics at arbitrary high energies. It may happen that new insights are needed and/or that some of the starting principles need to be slightly modified.\footnote{Another often-advocated possibility is that the QFT framework should be replaced by a different one to consistently describe physics at energies above a certain scale. On the contrary, we will provide both theoretical and phenomenological arguments that there is currently no indication that QFT and the strict renormalizability paradigm should be abandoned.} However, one of the messages we wish to convey through this paper is that, even if the paradigm of strict renormalizability will not turn out to be the final step in the evolutionary process of QFT, we cannot proceed without it and so the gravitational interaction must also conform to it.

In the next two sections we will describe gravity in the framework of perturbative local QFT, focus on the sub-Planckian regime and provide concrete evidence to support the last statement in the previous paragraph. Then, in Sec.~\ref{sec:Planck} we will comment on the (super-)Planckian regime. 

\section{Quantum gravity as a QFT}

After the SM was successfully formulated in the language of QFT, it became very natural to ask whether quantum aspects of gravity could be described in the same framework at the fundamental level. However, it was soon realized that general relativity (GR) is perturbatively non-renormalizable~\cite{tHooft:1974toh,Goroff:1985sz}, which  can be easily understood as follows. 

Consider the Einstein-Hilbert action
\begin{equation}
	S_{\rm EH}=\frac{1}{2\kappa^2}\int {\rm d}^4x\sqrt{-g}\left(R-2\Lambda_{\rm C}\right)\,,\label{action-GR}
\end{equation}
where $\kappa^2=8\pi G_{\rm N}=1/M_{\rm p}^2\,,$  $G_{\rm N}$ is Newton's constant, $M_{\rm p}\simeq 2.4\times 10^{18}$ GeV is the (reduced) Planck mass, and  $\Lambda_{\rm C}\simeq 10^{-122}M_{\rm p}^2$ is the cosmological constant.
Expanding in terms of a canonically normalized metric fluctuation $h_{\mu\nu}$ around some fixed background $\bar{g}_{\mu\nu}$, i.e. $g_{\mu\nu}=\bar{g}_{\mu\nu}+2\kappa h_{\mu\nu},$ we can note that $\kappa=1/M_{\rm p}$ and its positive powers play the role of interaction couplings of negative mass dimension. Additionally, it can be verified that the superficial degree of divergence of the most divergent diagrams is equal to $2L+2$, where $L$ is the number of loops: the UV divergences get worse as the number of loops increases. These aspects indicate that GR is non-renormalizable when quantized in the framework of perturbative QFT.

Despite this failure, a well-defined effective field theory (EFT) treatment of GR can still be formulated~\cite{Donoghue:1994dn} and quantum-gravity predictions can be trusted up to energies below a cutoff scale which, for example, in pure Einstein's gravity is given by the Planck mass but can also be lower if the coupling to matter is switched on~\cite{Han:2004wt,Dvali:2007hz,Dvali:2007wp}. The local part of this gravitational EFT action is given by
\begin{equation}\label{EFT-action}
\begin{aligned}
S_{\rm EFT}=\int {\rm d}^4x\sqrt{-g}\bigg[&\frac{1}{2\kappa^2}\left(R-2\Lambda_{\rm C}\right) +  a_1 R^2 + a_2 R_{\mu\nu}R^{\mu\nu} \\
&+a_3 \kappa^2 R^3+a_4\kappa^2 R_{\mu\nu\rho\sigma}R^{\rho\sigma}_{\phantom{\rho\sigma}\alpha\beta}R^{\alpha\beta\mu\nu}+\cdots\bigg]\,,
\end{aligned}
\end{equation}
where the dots stand for all possible local operators that are compatible with the geometric structure and symmetries of GR. In the absence of matter, the terms that are proportional to the classical field equations can be removed by performing a field redefinition of the metric tensor~\cite{Basile:2024oms}. 
The coefficients $a_i$ are dimensionless and their physical value can be obtained by renormalization up to errors proportional to positive powers of the ratio $E/M_{\rm p}$, where $E$ is some characteristic energy scale and $M_{\rm p}$ could also be replaced by a lower cutoff if matter is present. It is important to remark that in this EFT description the higher-order operators do not introduce any extra degrees of freedom in addition to the massless spin-two graviton.

The story that is usually told is that the perturbative QFT framework starts failing to provide an accurate description of the gravitational interaction at energy scales where the EFT of GR breaks down and that one should therefore resort to some non-perturbative QFT methods or opt for a beyond-QFT approach to analyse these and higher energy scales~\cite{Rocci:2024vrq,WALLACE202231,Bonanno:2020bil,Loll:2019rdj,Surya:2019ndm}. 

However, it is certainly good methodological practice, before abandoning the perturbative QFT framework for quantum gravity, to seek a four-dimensional QFT of gravity that extends GR at high energies and is compatible with the guiding principles listed in the previous section. In particular, we should ask the following question:
\begin{equation}
\centering \text{Does a strictly renormalizable QFT of gravitational interaction exist?}\nonumber
\end{equation}
We will now show that the answer is YES and then analyse the theoretical and phenomenological implications of the strict renormalizability paradigm for quantum gravity.

\subsection{Strict renormalizability paradigm: quadratic gravity}

This ``gravitational'' challenge is very similar in spirit to that of the electroweak interaction that we briefly discussed in the previous section. Does the failure of perturbative renormalizability of Fermi's theory imply that high-energy aspects of the weak interaction cannot be described within perturbative QFT? Of course the answer is NO, since we know that there exists a strictly renormalizable perturbative completion given by the Glashow-Weinberg-Salam electroweak theory~\cite{Glashow:1961tr,Weinberg:1967tq,Salam:1968rm}. 

Remarkably, a strongly analogous positive answer can be given for the gravitational interaction. In fact, in four spacetime dimensions there exists a \textit{unique} strictly renormalizable QFT of gravity that recovers GR in some low-energy regime and that preserves the geometric structure (metric compatibility and zero torsion) and the symmetries (diffeomorphism and parity) of GR. This theory is known as \textit{quadratic gravity}~\cite{Stelle:1976gc,Tomboulis:1977jk,Julve:1978xn,Fradkin:1981iu,Avramidi:1985ki,Salam:1978fd,Tomboulis:1980bs,Tomboulis:1983sw,Salvio:2016vxi,Salvio:2019ewf,Salvio:2018kwh, Salvio:2018crh,Salvio:2020axm,Salvio:2024joi,Holdom:2015kbf, Holdom:2021hlo,Holdom:2021oii,Holdom:2023usn,Holdom:2024cfq,Anselmi:2017ygm,Anselmi:2018ibi,Anselmi:2018tmf,Anselmi:2018bra,Piva:2023bcf,Donoghue:2018izj,Donoghue:2019ecz,Donoghue:2019fcb,Donoghue:2021cza,Donoghue:2021meq} and its classical bare action contains all possible operators of mass dimension up to four:
\begin{equation}
S_{\rm QG}=\frac{1}{2}\int {\rm d}^4x\sqrt{-g}\left[\frac{1}{\kappa^2}\left(R-2\Lambda_{\rm C}\right) + \frac{c_0}{6} R^2 -\frac{c_2}{2}C_{\mu\nu\rho\sigma}C^{\mu\nu\rho\sigma}\right]\,,
\label{quad-gravity-action}
\end{equation}
where $c_0$, $c_2$ are dimensionless parameters and $C_{\mu\nu\rho\sigma}$ is the Weyl tensor. The full bare action also includes the surface term $\sqrt{-g}\,\Box R$ and the Gauss-Bonnet $\sqrt{-g} \, \left(R_{\mu\nu\rho\sigma}R^{\mu\nu\rho\sigma}-4 R_{\mu\nu}R^{\mu\nu}+R^2\right)$ which is topological in four dimensions, but we have neglected both of them since they are total derivatives. However, it should be noted that both terms are generated by loop corrections and are important for the renormalization of the theory.

\medskip

\textbf{Strict renormalizability.} The quadratic curvature terms introduce fourth-order derivatives in the field equations and fourth powers of the momentum in the propagator and vertices. In this case the superficial degree of divergence of the most divergent diagram is equal to four. This means that UV divergences in quadratic gravity do not become worse as the number of loops increases and so only counterterms having the same functional form as those in the bare action are needed to perturbatively renormalize the theory to any loop order~\cite{Stelle:1976gc}. 

Considering metric perturbations around Minkowski, i.e. $g_{\mu\nu}=\eta_{\mu\nu}+2 h_{\mu\nu}$, we can expand the action~\eqref{quad-gravity-action} as follows (from now on, the cosmological constant is assumed to be negligible) 
\begin{equation}\label{quad-expans}
\begin{aligned}
S_{\rm QG}[\eta,h]= \int {\rm d}^4 x \bigg[ &\frac{1}{2}h_{\mu\nu}\Box \left(\frac{1}{\kappa^2}-c_2\Box\right)h^{\mu\nu}-h_\mu^\rho \left(\frac{1}{\kappa^2}-c_2 \Box\right)\partial_\rho \partial_\nu h^{\mu\nu}  \\
& +h \left( \frac{1}{\kappa^2}-\frac{1}{3}(2c_0+c_2)\Box \right)\partial_\mu\partial_\nu h^{\mu\nu}-\frac{1}{2}h \left(\frac{1}{\kappa^2}-\frac{1}{3}(2c_0+c_2)\Box \right)\Box h\\
&   + \frac{1}{3}(c_0-c_2)h_{\mu\nu}\partial^\mu\partial^\nu\partial^\rho\partial^\sigma h_{\rho\sigma}  \bigg] +S_{\rm QG}^{(n\geq 3)}[\eta,h]\,,
\end{aligned}
\end{equation}
where $S_{\rm QG}^{(n\geq 3)}[\eta,h]$ contains different types of interaction terms which are schematically given by
\begin{eqnarray}
\frac{1}{\kappa^2}\partial^2 h^n\,, \qquad c_0 \partial^4 h^{n}\,,\qquad  c_2 \partial^4 h^{n}\,.
\end{eqnarray}
The presence of both second- and fourth-order derivatives imply that the field $h_{\mu\nu}$ can be chosen with mass dimension equal to one or zero. In fact, the former choice is suitable in the low-energy regime where GR is dominant and an EFT treatment is still reliable, while the latter is the appropriate one at high energies. The possibility of having a zero-dimensional field is one of the reasons why the UV behavior of quadratic gravity is very different from that of GR and its EFT.

By performing a power-counting analysis of loop diagrams, we can easily verify that the UV behavior of the perturbative expansion in quadratic gravity is controlled by \textit{inverse} powers of the quadratic curvature coefficients, i.e. by the interaction couplings
\begin{equation}\label{inverse-coupl}
g_0\equiv \frac{1}{c_0}\,,\qquad g_2\equiv \frac{1}{c_2}\,.
\end{equation}
This fact implies at least two things: the interaction couplings governing the UV behavior are dimensionless, which is why the theory is \textit{strictly} renormalizable; the larger $c_0$ and $c_2$ are, the better the perturbative QFT expansion behaves~\cite{Basile:2024oms}.

When couplings to SM matter fields are switched on, we can still obtain a strictly renormalizable QFT if we also introduce a non-minimal coupling between the Higgs field and the metric given by the term $\int{\rm d}^4x\sqrt{-g} \, \xi \, |H|^2R
$~\cite{Stelle:1976gc,Salvio:2018crh}\,, where the interaction coupling $\xi$ is dimensionless. In addition to the free parameters of the SM, the extra couplings required by the strict renormalizability paradigm applied to the coupled gravity-matter system are finite in number and given by $G_{\rm N}$, $\Lambda_{\rm C}$, $c_0$, $c_2$ and $\xi$. 

\medskip

\textbf{Propagator and degrees of freedom.} The quadratic curvature operators are also responsible for the introduction of new massive degrees of freedom in addition to the massless spin-two graviton. This can be easily understood by looking at the spin structure and the poles of the (saturated) propagator~\cite{Stelle:1976gc}:
\begin{equation}\label{propag-spin-proj-quad-gra}
\begin{aligned}
\mathcal{G}_{\mu\nu\rho\sigma}(p^2)&= -i\Bigg(\frac{m_2^2 \mathcal{P}^{(2)}_{\phantom{(2)}\mu\nu\rho\sigma}}{p^2(p^2+m_2^2)}-\frac{m_0^2 \mathcal{P}^{(0,s)}_{\phantom{(0,s)}\mu\nu\rho\sigma}}{2p^2(p^2+m_0^2)}\Bigg)\\
&=-\frac{i}{p^2}\left(\mathcal{P}^{(2)}_{\phantom{(2)}\mu\nu\rho\sigma}-\frac{1}{2}\mathcal{P}^{(0,s)}_{\phantom{(0,s)}\mu\nu\rho\sigma}\right)-\frac{i}{2}\frac{\mathcal{P}^{(0,s)}_{\phantom{(0,s)}\mu\nu\rho\sigma}}{p^2+m_0^2}+i\frac{\mathcal{P}^{(2)}_{\phantom{(2)}\mu\nu\rho\sigma}}{p^2+m^2_2}\,,
\end{aligned}
\end{equation}
where the operators $\mathcal{P}^{(2)}$ and $\mathcal{P}^{(0,s)}$ are spin projectors whose explicit expressions are not important for our purposes (see~\cite{Basile:2024oms} for details). The contribution in the parentheses in the second line corresponds to the GR massless pole, while the second term represents a massive spin-zero mode coming from $R^2$, and the last term a massive spin-two ghost mode coming from $C_{\mu\nu\rho\sigma}C^{\mu\nu\rho\sigma}.$ On-shell, the massless spin-zero pole and the $0,\pm 1$ helicities of the massless spin-two pole do not contribute due to gauge invariance, so in total we have $2+1+5=8$ dynamical degrees of~freedom.

The masses squared of the additional degrees of freedom are given by\footnote{It is worth mentioning that the mass of the spin-two ghost depends on the cosmological constant $\Lambda_{\rm C}$, indeed the more precise expression (in four spacetime dimensions) is $m^2_2=\frac{M^2_{\rm p}}{c_2}+\frac{2}{3}\Lambda_{\rm C}\big(2\frac{c_0}{c_2}+1  \big)$~\cite{Anselmi:2018tmf,Buoninfante:2023ryt}.}
\begin{equation}\label{masses-spin-0-2}
m_0^2\equiv \frac{M_{\rm p}^2}{c_0}\,,\qquad m_2^2\equiv\frac{M_{\rm p}^2}{c_2}\,,
\end{equation}
where $c_0$ and $c_2$ must be positive in order to avoid tachyons. Note that,  $c_0,c_2\gg 1$ gives $m_0,m_2\ll M_{\rm p},$ which imply the existence of new physics beyond GR in the sub-Planckian regime. In fact, as already mentioned, the quadratic curvature coefficients are expected to be large in order to have a well-behaved QFT perturbative expansion. This feature can have extremely important phenomenological implications, as we will show in Sec.~\ref{sec:nature}.

\medskip

\textbf{Running couplings.} Unlike Newton's constant and the cosmological constant, which are not characterized by a physical running~\cite{Anber:2011ut,Donoghue:2024uay}, the quadratic curvature coefficients run as a function of the physical energy or momentum. First computations of beta functions in quadratic gravity trace back to forty years ago~\cite{Julve:1978xn, Fradkin:1981iu, Avramidi:1985ki}. However, it has been recently claimed by~\cite{Buccio:2024hys} that these old results do not capture the true physical running of the couplings due to the fact that tadpoles (that do not carry any physical momentum dependence) were inappropriately included as contributions in the one-loop beta functions and infrared terms (that can be relevant in four-derivative theories) were not taken into account. It is believed that further investigations are still needed to definitively settle the question of what the physical gauge-independent beta functions are for the quadratic curvature coefficients.

It is worth mentioning that both old and new calculations show the existence of asymptotically free solutions for the couplings $g_0$ and $g_2.$ Despite the fact that these couplings can flow to zero in the infinite energy limit, it can be shown that the exclusive cross sections in quadratic gravity still grow as a function of the energy~\cite{Holdom:2021hlo}. In other words, the suppressed behavior of the interaction couplings seems not to be sufficient to obtain a suppressed behavior of the amplitudes at high energies. This might suggest that the concept of asymptotic freedom and the meaning of the UV limit need to be rethought in a quantum-gravitational context; see Sec.~\ref{sec:Planck} for further discussion.

\medskip

All these theoretical aspects that we have briefly discussed
completely distinguish quadratic gravity from the EFT of GR. However, some skeptics may still wonder why these two theories are so different if they both contain quadratic curvature operators in their actions (see Eqs.~\eqref{quad-gravity-action} and~\eqref{EFT-action}). Indeed, this question and its answer are often a source of confusion and misunderstanding. Therefore, to better grasp the key features of quadratic gravity and the main differences with the EFT of GR, we will now make an instructive comparison between the two theories.

\subsection{Quadratic gravity vs EFT of general relativity}

Let us imagine that we are interested in some physical process whose characteristic energy scale $E$ is much lower than the EFT cutoff, e.g. the Planck mass in pure Einstein's gravity. Furthermore, consider a quadratic curvature truncation of the EFT expansion~\eqref{EFT-action}, so that any EFT prediction can be made up to an accuracy of the order of $(E/M_{\rm p})^4.$ Up to boundary terms, this truncated EFT action has the same structure as that of quadratic gravity in~\eqref{quad-gravity-action}. Then, an obvious question to ask could be: Do these two actions describe the same physics in the sub-Planckian regime, i.e. for energies $E\ll M_{\rm p}$? The answer is NO, as we will now explain in detail.

\medskip

\textbf{Different perturbative expansions.} In the EFT of GR the kinetic term is of second order in the derivatives, while the one in quadratic gravity contains both second and fourth-order derivatives acting on $h_{\mu\nu}$. This means that, in the path integral language, the perturbative expansions in the two theories are performed around two different Gaussians. While the propagator in the EFT of GR falls off like $1/p^2,$ the one in quadratic gravity is more suppressed and behaves as $1/p^4.$  These are the reasons why in quadratic gravity we can work with a dimension-zero field and then prove perturbative renormalizability. It is also worth mentioning that, starting from quadratic gravity, we can define an EFT expansion if we are interested in energy scales lower than $m_0$ and/or $m_2$, by integrating out the massive spin-zero and/or spin-two degrees of freedom.

\medskip
    
\textbf{Different meanings of quadratic curvature coefficients.} In a perturbative regime, $a_1$ and $a_2$ appearing in the EFT action~\eqref{EFT-action} are expected to have renormalized values of order one or smaller. On the contrary, the coefficients $c_0$ and $c_2$ in the quadratic gravity action~\eqref{quad-gravity-action} are free parameters that can be renormalized to arbitrary values, even large ones, provided that the perturbative expansion can be considered reliable. 

\medskip

\textbf{Inessential vs essential couplings.} In the absence of matter, the quadratic curvature terms in the EFT of GR can be removed perturbatively by performing a metric field redefinition (the same is true for any other term that is proportional to the Einstein's equations in vacuum). If matter is present, their contributions can still be partially removed by shifting them into the matter sector. On the other hand, in the renormalizable theory of quadratic gravity this is not possible: $R^2$ and $C_{\mu\nu\rho\sigma}C^{\mu\nu\rho\sigma}$ are part of the starting action, they lead to field equations different from those of GR and therefore to a different dynamics for the metric field in high-curvature regimes. In this case, the same field redefinition would have a different effect since it would map the action of quadratic gravity to an equivalent action containing higher curvature operators that cannot be neglected perturbatively. In other words, while the EFT coefficients $a_1$ and $a_2$ are inessential couplings, the quadratic gravity coefficients $c_0$ and $c_2$ are essential.

\medskip
    
\textbf{Small vs large couplings.} While in the EFT of GR the loop diagrams are proportional to $(1/M_{\rm p}^2)^{L-1}$ and positive powers of $a_i$, in quadratic gravity the UV behavior of the loop expansion is controlled by $(1/c_0)^{L-1}$, $(1/c_2)^{L-1}$, and other dimensionless combinations (see also~\cite{Basile:2024oms}). This means that what is strongly (weakly) coupled from the point of view of quadratic gravity is weakly (strongly) coupled from the point of the EFT of GR. Note that, large values of $a_i$ correspond to a strong coupling regime for the EFT only if the corresponding energy scale is smaller than the EFT cutoff.  However, if the EFT cutoff is sufficiently low, for example much lower than the Planck mass, then hitting high values of $a_i$ could simply be a sign of the presence of new physics that has not been properly accounted for. A clear example of this latter scenario is in fact given by quadratic gravity taken as  perturbative completion of GR, according to which the EFT cutoff corresponds to the smallest of the two masses $m_0$ and $m_2$. In this case, the gravitational interaction would still be weakly coupled for energy scales $E\sim m_i$, and even larger.

\medskip

\textbf{Different number of degrees of freedom.} If the EFT cutoff is given by the Planck mass (or slightly lower due to coupling with matter), then the EFT of GR does not predict extra degrees of freedom in addition to the massless spin-two graviton at energy scales $E\ll M_{\rm p}$. In contrast, quadratic gravity predicts the existence of new degrees of freedom that can be active already at energy scales of the order of their masses, i.e. $m_0=M_{\rm p}/\sqrt{c_0}$ and $m_2=M_{\rm p}/\sqrt{c_2}$, whose values are sub-Planckian since the coefficients $c_0$ and $c_2$ are required to be large in order to have a well-behaved perturbative expansion in quadratic gravity. The presence of additional massive degrees of freedom is one reason why quadratic gravity represents a perturbative QFT-based completion of GR.
    
\medskip

From this comparison between the EFT of GR and quadratic gravity, it should now be clear that the latter can describe new physics in the sub-Planckian regime, e.g. in the early universe, as we will explain in Sec.~\ref{sec:nature}. Furthermore, the above discussion is also useful to contrast the old view of strict renormalizability as a selection principle with the modern view that all QFTs, including the strictly renormalizable ones, are just EFTs~\cite{Weinberg:1978kz,Weinberg:1996kw,Weinberg:2009bg,Rocci:2024vrq}; see also the end of Sec.~\ref{sec:history}. 

\subsection{Uniqueness}

The criterion of strict renormalizability avoids the proliferation of higher-order operators in the bare Lagrangian. Having a finite number of parameters makes the theory much more predictive and falsifiable. It is important to emphasize again that in four spacetime dimensions quadratic gravity is a \textit{unique} strictly renormalizable gravitational QFT~\cite{Stelle:1976gc,Basile:2024oms}, which is expected to recover GR in the low-energy regime (i.e. for energies lower than $m_0$ and $m_2$), is metric compatible, has zero torsion and respects the symmetries of GR. This means that if future experiments or observations falsify quadratic gravity, we will then be forced to modify some of the starting~principles.

It is worth mentioning that if we remove the Einstein-Hilbert term from~\eqref{quad-gravity-action}, we are left with a purely quadratic Lagrangian that is still strictly renormalizable. However, in this case GR cannot be recovered at low energies, unless the Einstein-Hilbert term is induced via a matter coupling~\cite{Salvio:2014soa}. If we remove the $R^2$ term as well, we are left with the Lagrangian of conformal gravity (in four dimensions)~\cite{Mannheim:2011ds,tHooft:2015tlq,Jizba:2014taa}, which is non-renormalizable from two loops~\cite{Fradkin:1983tg,Salvio:2017qkx}. Furthermore, if we introduce non-metricity and a non-zero torsion we might still achieve strict renormalizability~\cite{Percacci:2023rbo,Melichev:2023lwj}. Similarly, if we break local Lorentz invariance we might still hope to find renormalizable Lagrangians~\cite{Herrero-Valea:2023zex}. However, as soon as we depart from the geometric and symmetry properties of GR, we open up a wider class of possibilities and lose uniqueness. From this point of view, quadratic gravity is the most conservative quantum theory of gravity we can~conceive.

This uniqueness property excludes also super-renormalizable and/or non-local QFTs of gravity~\cite{Anselmi:2018cau}. In fact, if we admit the presence of derivatives of order higher than four or of non-polynomial differential operators, there exists an infinite number of gravitational bare Lagrangians that are either local and super-renormalizable~\cite{Asorey:1996hz,Modesto:2015ozb,Anselmi:2017ygm} or non-local and renormalizable~\cite{Modesto:2011kw,Koshelev:2017ebj,BasiBeneito:2022wux,Buoninfante:2021xxl}. This means that if an experiment falsifies one of these Lagrangians, then we can always choose another one that respects the same starting principles. Therefore, there is no way to actually falsify the assumptions of super-renormalizability and non-locality for the bare Lagrangian.

\subsection{What about the massive spin-two ghost?}

A feature of quadratic gravity that has raised concerns in the past and stimulated new ideas in recent years~\cite{Buoninfante:2022ykf,Buoninfante:2024yth} is the ghostly nature of the massive spin-two field arising from the Weyl-squared term. Looking at the propagator~\eqref{propag-spin-proj-quad-gra}, we can see that the sign of the last component is opposite to the standard case, due to unconstrained fourth-order derivatives acting on the metric.
Ghosts could cause Hamiltonian instabilities due to the Ostrogradsky theorem~\cite{Woodard:2015zca} and violate unitarity~\cite{Basile:2024oms}. However, these statements rely on specific assumptions and there are~loopholes. 

First, Ostrogradsky theorem applies only to Lagrangians that are non-degenerate, but this is not the case for gravitational Lagrangians due to diffeomorphism invariance. Therefore, strictly speaking the theorem does not apply to quadratic gravity. Indeed, numerical relativity analyses have shown that non-linear evolutions of perturbations in spacetimes with single or binary black holes can be stable~\cite{Held:2023aap,Held:2025ckb}. Moreover, stable classical solutions have been found in some toy models with ghosts~\cite{Deffayet:2021nnt,Deffayet:2023wdg,ErrastiDiez:2024hfq}. The possibility of physically viable metastable scenarios have also been considered~\cite{Salvio:2019ewf,Gross:2020tph}. Furthermore, the question of classical instabilities might not arise at all if the quantum theory were stable~\cite{Anselmi:2018bra,Donoghue:2021eto}. This is not the end of the story, but there is certainly no proof implying that quadratic gravity is physically inconsistent due to classical instabilities.

Second, the claim about unitarity violation relies heavily on the simultaneous assumptions of positive (squared) norms for ghost states and Feynman prescription ($p^2\rightarrow p^2-i\epsilon$) for the ghost propagator. However, unitarity can be shown to hold if alternative quantization prescriptions are implemented. There are at least three approaches~\cite{Basile:2024oms}: (\textit{i}) choosing negative norms for states containing an odd number of ghost particles and retaining the Feynman prescription for all propagators~\cite{Stelle:1976gc,Salvio:2018kwh, Salvio:2018crh,Salvio:2020axm,Salvio:2024joi,Holdom:2015kbf, Holdom:2021hlo,Holdom:2021oii,Holdom:2023usn,Holdom:2024cfq}; (\textit{ii}) retaining positive norms for all physical states and prescribing the ghost propagator with the anti-Feynman shift together with new rules to compute loop integrals~\cite{Donoghue:2019ecz,Donoghue:2019fcb,Donoghue:2021cza}; (\textit{iii}) converting the ghost into a purely virtual particle  (fakeon) by prescribing the ghost propagator with an average of Feynman and anti-Feynman prescriptions together with new rules to compute loop integrals~\cite{Anselmi:2017ygm,Anselmi:2018ibi,Anselmi:2018tmf,Anselmi:2018bra,Piva:2023bcf}. 

In our opinion, these unitary quantizations still face some open questions (see~\cite{Basile:2024oms}). Here we will limit ourselves to making a few remarks and refer readers to the original works for details. 

Unlike the first prescription for which the standard Wick rotation still applies, the second and third require alternative contour deformations for the evaluation of the loop integrals. In the latter cases the theory is formulated at the level of Feynman diagrams by new diagrammatic rules. 
Of the three, only the first seems compatible with the  operator formalism of local~QFT. In this case, the ghost states have positive energy and the (quantum) Hamiltonian is positive-definite. However, it is not yet clear whether negative norms give rise to observable negative probabilities~\cite{Woodard:2023tgb,Kubo:2023lpz}. An alternative QFT formulation (e.g. with a pseudo-Hermitian Hamiltonian~\cite{Salvio:2015gsi,Strumia:2017dvt,Kuntz:2024rzu}) and/or a modified Born rule~\cite{Salvio:2020axm,Holdom:2024onr} may be needed, or  it could simply be that cancellations eventually occur and the physical observables are those associated with positive probabilities~\cite{Feynman1987-FEYNP,Holdom:2021hlo,Holdom:2021oii,Holdom:2023usn}. 

A striking feature of ghost propagators is that, even if they acquire a non-zero width at the quantum level, the corresponding ghost particles cannot decay (or, at least, not in the standard way). Indeed, it can be shown that the pair of complex conjugate poles does not lie in the second Riemann sheet of the complex plane as for ordinary unstable particles, but appears in the first sheet~\cite{Coleman:1969xz,Kubo:2024ysu,Buoninfante:2025klm}. According to the operator formalism of QFT, this should imply that ghosts remain part of the set of asymptotic states.
In our opinion, acquiring a better understanding of ghost resonances is crucial to learn more about quadratic gravity~\cite{Buoninfante:2025klm}. In particular, we have recently pointed out that finite-time QFT may be needed to consistently define these objects and capture their true physical features~\cite{Buoninfante:2025klm,Anselmi:2023wjx}.

The presence of the spin-two ghost may induce the violation of some notion of causality in the high-energy regime, i.e. at time scales of the order of or smaller than $\mathcal{O}(M_{\rm p}^2/m_2^3)$~\cite{Coleman:1969xz,Anselmi:2018tmf,Donoghue:2019ecz,Donoghue:2021meq}. 
This means that the usual notion of causality may be an emergent phenomenon~\cite{Grinstein:2008bg}. Furthermore, the correspondence principle may not hold in quadratic gravity and it may be necessary to define the classical theory as a consistent limit of the corresponding quantum theory~\cite{Anselmi:2018bra,Donoghue:2021eto}. For example, the ghost quantized as a fakeon does not appear as an on-shell state since it is projected out of the physical Hilbert space by the new diagrammatics~\cite{Anselmi:2021hab}. In this case, the action of quadratic gravity~\eqref{quad-gravity-action} is an ``interim'' action, while the true classical action is non-local~\cite{Anselmi:2018bra}. 

How the violation of causality manifests itself and the details of the classical limit may depend on the type of quantization prescription. This also means that different quantization prescriptions can lead to different physical implications.

Furthermore, the known quantizations of quadratic gravity are performed perturbatively, i.e. the theory is formulated in the framework of perturbative QFT. It is not yet clear whether all three quantizations mentioned above or only some of them can be derived from a non-perturbative formulation of quadratic gravity. See~\cite{Salvio:2024joi} for some recent progress.

The exciting part of all this is that there is still more to learn about the conceptual aspects and physical implications of QFT and gravity. Skeptics may be hesitant to implement alternative quantizations for ghost fields and prefer to abandon quadratic gravity and with it strict renormalizability. However, it may not be surprising that a non-standard quantization is ultimately necessary. We already know that the quantization rules for Dirac fields are different from those for Klein-Gordon fields. It is true that in this particular example the different statistics also play an important role, but our point is that it may indeed be that the type of quantization depends on the number of derivatives~\cite{Salvio:2015gsi}.
Furthermore, and more importantly: in the next section we will show that Nature continues to favor strict renormalizability as a paradigm for fundamental physics, which should further motivate us to take quadratic gravity and ghosts more seriously.

\section{Let Nature guide us}\label{sec:nature}

It is often argued that, due to the lack of experimental and observational data, we cannot use experiments and observations to guide us towards a theory of quantum gravity that goes beyond GR and, therefore, we can only resort to theoretical consistency requirements~\cite{Rickles:2020zwh,Rocci:2024vrq,Basile:2025zjc} or numerical tools~\cite{Loll:2022ibq}. Instead, we will now provide solid, concrete evidence showing that we can actually use guidance from~observations.

The following \textit{two} steps are quite logical and scientifically reasonable:
\begin{enumerate}[label=(\textit{\roman*})]

    \item If new experiments or observations cannot be explained by pure GR, then we can use them to guide us toward understanding what kind of additional terms need to be added to the Einstein-Hilbert action to describe the observed new physics;

    \item After having phenomenologically modeled the new terms, we ask whether there is any fundamental guiding principle or mechanism that can justify the new Lagrangian.

\end{enumerate}
For this second step, it is very natural to first check whether what has worked in the past as a selection principle can be useful for gravity as well, in particular whether the QFT framework and the strict renormalizability criterion can successfully select a gravitational Lagrangian through which the new physics can be described.

\subsection{Early universe cosmology and $R^2$-driven inflation}

It is well-known that the Einstein-Hilbert action \textit{alone} fails to provide a satisfactory physical description of our universe at early times, in particular it cannot explain the observed temperature anisotropies in the Cosmic Microwave Background (CMB)~\cite{Planck:2018jri,Planck:2018vyg}. 

A simple way to interpret the CMB data is to assume that at very early times there was some sort of accelerated expansion phase, driven by some field whose subsequent quantum fluctuations would have acted as inhomogeneous seeds for the formation of large-scale structures and galaxies, thus leaving an imprint in the CMB in the form of temperature anisotropies.

Phenomenologically, this primordial inflationary phase is usually described by adding by hand a scalar field $\phi$ coupled to gravity and with a suitable self-potential, i.e. 
\begin{equation}
	S_{\rm EH}[g]\quad \rightarrow \quad S_{\rm EH}[g]+ S_{\phi}[g,\phi]\,,
\end{equation}
where $S_\phi$ is the contribution of the self-interacting scalar field coupled to gravity. Some of the free parameters appearing in $S_\phi$ can then be fitted by matching with CMB observations. Of course, to have a consistent picture of our universe we should also add other contributions due to ordinary and dark matter, in addition to the cosmological constant one.\footnote{Here we are not concerned  with the problems of dark matter and dark energy. In any case, we are comfortable with having a small cosmological constant which, in quadratic gravity, is simply a free parameter that can be renormalized to any measured value.}

Now, from a theoretical point of view, the natural question to ask is how to justify the additional term in the action at a fundamental level. One possibility is to think of $S_{\phi}$ as simply given by a new matter field. For instance, (infinitely) many models of inflation have been proposed where the inflaton field is a new matter degree of freedom with some ad-hoc chosen self-potential~\cite{Martin:2013tda}. However, this way of proceeding would then raise the further question of how to choose the new matter field and, in particular, how to \textit{uniquely} select the shape of the self-potential. Instead, to implement step (\textit{ii}) above, we now ask whether the strict renormalizability criterion
can help us to obtain a gravitational Lagrangian that naturally incorporates a term like $S_\phi.$

We already know that if we work in the QFT framework and stay as conservative as possible, then we end up selecting the strictly renormalizable action of quadratic gravity~\eqref{quad-gravity-action}. A key feature is that the $R^2$ term introduces a scalar degree of freedom with mass squared $m_0^2=M_{\rm p}^2/c_0,$ known as the \textit{scalaron}. This can also be seen by introducing an auxiliary scalar field $\phi$ and making a conformal metric transformation to go to the Einstein frame, in which the action~\eqref{quad-gravity-action} reads~\cite{Anselmi:2018tmf,Buoninfante:2023ryt}
\begin{equation}
S_{\rm QG}[g,\phi]=S_{\rm EH}[g]+S_\phi[g,\phi]-\frac{c_2}{4}\int {\rm d}^4x\sqrt{-g}C_{\mu\nu\rho\sigma}C^{\mu\nu\rho\sigma}\,,
\label{action-einstein-frame}
\end{equation}
where  
\begin{equation}
S_\phi[g,\phi]=\int {\rm d}^4x\sqrt{-g}\left[ -\frac{1}{2}\nabla_{\mu}\phi\nabla^\mu\phi -\frac{3m_0^2}{4\kappa^2}\left(1-e^{-\sqrt{2/3}\,\kappa\phi}\right)^2 \right]\,;\label{action-phi}
\end{equation}
for simplicity we have neglected the cosmological constant and with an abuse of notation we have denoted the transformed metric with the same symbol.

The self-potential has an infinite plateau, which is actually very suitable for a consistent description of the inflationary phase. In fact, if we drop the Weyl-squared term from~\eqref{quad-gravity-action} (or~\eqref{action-einstein-frame}) we obtain Starobinsky's model of inflation~\cite{Starobinsky:1980te,Mukhanov:1981xt,Starobinsky:1981vz,Starobinsky:1983zz,Kofman:1985aw,Vilenkin:1985md},\footnote{Starobinsky inflation was initially discussed as a consequence of one-loop trace anomaly in the context of semi-classical gravity coupled to conformal matter fields~\cite{Starobinsky:1980te,Mukhanov:1981xt,Starobinsky:1981vz,Starobinsky:1983zz}. However, soon after it was realized that the model could be more easily and consistently formulated by starting from the action $R+R^2$~\cite{Kofman:1985aw,Vilenkin:1985md}.} which is one the models most favored by CMB data~\cite{Planck:2018jri,Ellis:2023wic}. Let us briefly review the merits of Starobinsky inflation; then in the next subsection we discuss the additional effects of the Weyl-squared term.

The measurement of the spectral index of the scalar power spectrum, whose leading-order expression in the slow-roll approximation is $n_s= 1-2/N_e$~\cite{Mukhanov:1981xt}, constrains the number of e-folds $N_e.$ Through the measurement of the amplitude, $A_s=m_0^2N_e^2/(24\pi^2M_{\rm p}^2)$~\cite{Starobinsky:1983zz,Ketov:2025nkr}, we can then fit the value of $m_0$.
The Planck data~\cite{Planck:2018vyg} give\footnote{The ACT collaboration~\cite{ACT:2025fju,ACT:2025tim} has recently announced a new measured value of the scalar spectral index, $n_s\simeq 0.9666\pm 0.0077$, which is consistent with Planck. They also claim that the combination of ACT and Planck, or ACT, Planck and DESI~\cite{DESI:2024uvr,DESI:2024mwx} datasets gives $n_s\simeq 0.974\pm 0.003$. This would imply $N_e\simeq 77\pm 9$ for Starobinsky inflation.
Further analyses of the combined datasets are needed before drawing definitive conclusions. In fact, another recent result from the combined datasets SPT+Planck~\cite{SPT-3G:2024atg} gives $n_s\simeq 0.9647 \pm 0.0037$, which is consistent with Planck. Furthermore, WMAP+SPT~\cite{SPT-3G:2024atg} and WMAP+ACT~\cite{DESI:2024uvr} are also compatible with Planck. We expect that eventually all the combined datasets will agree with Planck.}  $n_s\simeq 0.9649 \pm 0.0042$ and~$A_s\simeq (2.10\pm 0.03)\times10^{-9},$ thus we get $N_e\simeq 57\pm 7$ and the following value of the scalaron mass~(for $N_e=57$):
\begin{equation}
m_0\simeq 3.0\times 10^{13}\,\text{GeV}\qquad \Leftrightarrow\qquad c_0 \simeq 6.5 \times 10^{9}\,,\label{value-scalaron-mass}
\end{equation}
which implies that the energy scale of Starobinsky inflation is $(3/4)^{\frac{1}{4}}\sqrt{m_0M_{\rm p}}\simeq 7.9\times 10^{15}\,\text{GeV}.$ Here the values of $n_s$ and $N_e$ are defined at the horizon crossing with the pivot scale~$0.05\,\text{Mpc}^{-1}$.

It is important to mention that Starobinsky worked in the Jordan frame, i.e. directly with the action $R+R^2$. He showed that the corresponding field equations admit a quasi de Sitter solution, through which the accelerated primordial expansion can be successfully explained. 
During this initial phase the scale-invariant term $c_0 R^2$ dominates over $M_{\rm p}^2 R$, thus explaining the \textit{nearly} scale-invariant power spectrum. Another feature that makes Starobinsky inflation unique is that it can also describe the reheating phase at the end of inflation through particle production induced by oscillations of the scalaron and hence of the scale factor, i.e. $a(t)\propto t^{2/3} [1+2/(3m_0t)\, \sin (m_0 t)]$~\cite{Starobinsky:1981vz,Starobinsky:1983zz}. The reheating temperature compatible with the number of SM degrees of freedom was found to be of the order of $T_{\rm reh}\sim \mathcal{O}(10^{9})\,\text{GeV}$~\cite{Starobinsky:1981vz,Gorbunov:2010bn,Jeong:2023zrv}, but the details of the (p)reheating are still under investigation~\cite{Gorbunov:2010bn,Jeong:2023zrv,del-Corral:2023apl,Garcia:2023tkk,Dorsch:2024nan}. Furthermore, at later times Starobinsky's model predicts the behavior $a(t)\propto t^{1/2}$~\cite{Starobinsky:1981vz}, which nicely connects to the radiation-dominated era during which the GR term $M_{\rm p}R$ becomes dominant and, if supplemented with some baryogenesis mechanism, can explain the subsequent~standard cosmological~evolution.

The surprising aspect we want to emphasize is that the QFT framework and the strict renormalizability paradigm give for free (without us asking) an additional scalar degree of freedom whose dynamics can provide a successful description of the early universe evolution. Moreover, inflation, or rather the accelerated primordial expansion, is not just an hypothesis but a consequence of strict renormalizability. It should be appreciated that Nature is guiding us toward the selection of a new theory that extends GR in the high-energy and high-curvature regimes. The steps (\textit{i}) and (\textit{ii}) above appear to be compatible with the strict renormalizability paradigm.

Starobinsky inflation is usually considered just a model, mainly because it is non-renormali\-\-zable. However, we now know that strict renormalizability can be achieved by adding the Weyl-squared term. Indeed, quadratic gravity can be seen as an UV completion of Starobinsky's model.

\subsection{Inflationary implications of the Weyl-squared term}

Since the Weyl tensor vanishes for conformally flat metrics, it does not affect the homogeneous and isotropic background solutions found in $R+R^2.$ However, it does influence the metric perturbations. At the lowest order in the slow-roll approximation, the scalar and tensor fluctuations are decoupled, meaning that the power spectrum of the primordial scalar fluctuations is still given by that of Starobinsky's model. However, the Weyl-squared term modifies the power spectrum of the tensor perturbations in a non-trivial way, giving potentially different predictions.

A relevant measurable quantity is the tensor-to-scalar ratio defined as the ratio of the tensor to the scalar power spectrum. Its leading-order expression in the slow-roll approximation is independent of the type of quantization used for the spin-two ghost and reads~\cite{Deruelle:2012xv,Ivanov:2016hcm,Salvio:2017xul,Anselmi:2020lpp,Salvio:2022mld,Kubo:2025jla}
\begin{equation}\label{tensor-to-scalar} 
r=\frac{24}{N_{e}^2}\frac{m_2^2}{m_0^2+2m_2^2}=\frac{24}{N_{e}^2}\frac{c_0}{c_2+2c_0}\,.
\end{equation}
In the limit $m_2/m_0\rightarrow \infty$ we recover Starobinsky's expression, i.e. $r_{\rm Starob}= 12/N_e^2$~\cite{Ketov:2025nkr}. If we take Planck mission's central value $N_e=57$, we get $r_{\rm Starob}\simeq 0.0037.$  However, if the Weyl-squared term is non-negligible we could have different values. Indeed, a future measurement of the tensor-to-scalar ratio will directly fix the value of $m_2$ (i.e. $c_2$). This means that quadratic gravity will then be able to provide sharp predictions for the other cosmological parameters such us the running of the scalar spectral index and the tensor tilt $n_t$. The latter, at leading order, satisfies the consistency relation $n_t\simeq -r/8$~\cite{Anselmi:2020lpp,Salvio:2022mld,Kubo:2025jla}.

Although $m_2$ (i.e. $c_2$) is a free parameter whose value cannot be predicted and must be fixed by observations, from a high-energy physics point of view we might expect the effects of the spin-two ghost to become important at energies of the order or higher than the scalaron mass, $m_2\gtrsim m_0$ (i.e. $c_2\lesssim c_0$). 

Taking the range $(m_0,\infty)$ as the possible allowed values for $m_2,$ then  quadratic gravity would give the following narrow window for the tensor-to-scalar ratio~\eqref{tensor-to-scalar}:
\begin{equation}\label{tensor-to-scalar-pred} 
\frac{8}{N_e^2}\lesssim r\lesssim \frac{12}{N_e^2}\,. \end{equation}
In particular, for $N_e=57$ we get $0.0025\lesssim r\lesssim 0.0037.$ 

Values of $m_2$ lower than $m_0$ have also been considered in the literature. For example, if the spin-two ghost is quantized as a fakeon, the consistency of the quantization prescription requires $m_2\geq m_0/4$, so the lower bound in~\eqref{tensor-to-scalar-pred} would be replaced by $4/(3N_e^2)$ which for $N_e=57$ gives  $0.0004$~\cite{Anselmi:2020lpp}. Even smaller values have been considered in~\cite{Salvio:2018crh,Salvio:2022mld}. A measured value of the tensor-to-scalar ratio less than $\mathcal{O}(1)/N_e^2$ would, in principle, not rule out the theory. However, if quadratic gravity is a consistent UV completion of Starobinsky's model, then we would expect the value of the tensor-to-scalar ratio to be close to Starobinsky's prediction.

The current observational constraint on the tensor-to-scalar ratio comes from the Planck mission and reads $r\lesssim 0.032\,\,(95\%\, \text{CL})$~\cite{BICEP:2021xfz,Tristram:2021tvh}. Interestingly, we are only one order of magnitude away from Starobinsky's prediction and our expectations for quadratic gravity. Actually, future observations~\cite{CORE:2016ymi,CMB-S4:2020lpa,Paoletti:2022kij,NASAPICO:2019thw,SimonsObservatory:2018koc,Euclid:2021qvm} will look for values of the order of $r\sim \mathcal{O}(10^{-3})$, which is exactly what is needed to test quadratic gravity as a UV completion of Starobinsky's model.

Some of the observational missions, such as LiteBird~\cite{Paoletti:2022kij}, are expected to be launched in the next ten years or so. Therefore, in the near future we will be able to say whether Nature favors quadratic gravity as a physically viable extension of GR. If confirmed, this would be a spectacular new success of the QFT framework and the strict renormalizability paradigm!

As mentioned in the previous section, some form of causality is expected to be violated for time intervals $\Delta t\lesssim M_{\rm p}^2/m_2^3\lesssim M_{\rm p}^2/m_0^3 \sim 10^{-28}$~s due to the presence of the spin-two ghost. This time scale may be much larger than the duration of inflation (about $10^{-36}$~s), so future investigations are certainly needed to see whether microscopic acausal effects could leave some kind of imprint on the CMB correlations~\cite{Tomboulis:1980bs,Buoninfante:2024yth}.

\section{Remarks on the (super-)Planckian regime}\label{sec:Planck}

Is quadratic gravity UV-complete in the sense that it can make predictions at arbitrarily high energies? This question is still open~\cite{Basile:2024oms,Buoninfante:2024yth}. 

Quadratic gravity differs greatly from standard renormalizable QFTs. Unlike the case of the electroweak theory where, for example, the addition of the Higgs boson also improves the UV behavior of the tree-level $W\text{-}W$ cross section, in the case of quadratic gravity the additional massive degrees of freedom do not improve the UV behavior of the tree-level graviton-graviton amplitude. The latter is equal to the GR expression and grows as a function of the center-of-mass energy, i.e. as $E_{\rm cm}^2/M_{\rm p}^2$~\cite{Dona:2015tra,Holdom:2021hlo}. Furthermore, unlike the case of QCD where the property of asymptotic freedom causes the scattering amplitudes to vanish in the infinite energy limit, in quadratic gravity it is not yet clear whether the same holds true. At least in renormalizable scalar models the presence of fourth-order derivatives was shown to introduce a power-law dependence that overcomes the one-loop logarithmic suppression of the asymptotically free couplings~\cite{Buccio:2023lzo}.

These peculiar features could have some physical explanations and implications. For example, asymptotic freedom may not be necessary in a quantum-gravitational context; or having asymptotically free couplings may just be crucial to ensure that the perturbative expansion is well-defined in the UV regime, but additional insights may be needed to have a suppressed high-energy behavior of the amplitudes when higher-order derivatives are present. Let us discuss at least three possibilities.
\begin{itemize}

\item \textbf{Non-perturbative resummation:} The growth of the tree-level graviton-graviton scattering amplitude might just be an indication that perturbation theory breaks down at energy scales of the order of $E_{\rm cm}\sim M_{\rm p}.$ This could mean that the resummation of the loop contributions is needed in order to achieve suppression of the amplitudes~\cite{Aydemir:2012nz,Piva:2023eaj}. Due to computational complexity, this has not yet been verified in quadratic~gravity.

\item \textbf{Totally inclusive cross sections:} It might be that in quadratic gravity the exclusive cross sections are not physically observable at sufficiently high energies, but the measurable ones are totally inclusive~\cite{Holdom:2021hlo,Holdom:2021oii,Holdom:2023usn,Holdom:2024cfq}. This was first suggested in~\cite{Holdom:2021hlo}, where it was shown that the totally inclusive cross section of all tree-level exclusive processes, involving graviton, scalaron and spin-two ghost, is suppressed in the high-energy limit. This would mean that at energies $E_{\rm cm}\gtrsim m_0,m_2$ the presence of the additional massive degrees of freedom cannot be neglected and all possible active processes must be included. This phenomenon only occurs if the spin-two ghost is quantized with negative norms, since the presence of negative cross sections is important for cancellations to take place.  This idea is still a matter of debate, but it must be said that such a  cancellation could also help explain why negative probabilities may ultimately not be observable in quadratic gravity.

\item \textbf{Black-hole dominance:} If black holes are present in a gravitational theory, then they are expected to dominate in some high-energy regime~\cite{tHooft:1987vrq,Giddings:2007qq,Dvali:2014ila,Addazi:2016ksu}. Classical black-hole solutions can be found in quadratic gravity~\cite{Lu:2015cqa,Lu:2015psa,Buoninfante:2024oyi} and so they could form via a scattering process (see~\cite{Holdom:2016nek,Holdom:2021hlo} for a different point of view). Indeed, the growth of the tree-level graviton-graviton amplitude might simply suggest that at high energies the non-perturbative phenomenon of black-hole formation occurs and we must somehow take this into account to get suppressed amplitudes. This would mean that renormalizability has to be supplemented by a built-in classicalization mechanism~\cite{Dvali:2014ila}, according to which the amplitude gets entropically suppressed as $e^{-E_{\rm cm}^2/M_{\rm p}^2}$ for energies above the threshold for black-hole formation.

\end{itemize}

These three options may not be mutually exclusive. Indeed, we may have a high-energy regime where the impact parameter is still large enough not to trigger black-hole formation, so that the role played by totally inclusive cross sections may be important. On the other hand, if we are in a high-energy regime where the impact parameter is small enough, then black-hole formation is triggered and prevails over all other possible processes. Furthermore, the non-perturbative resummation discussed in the first item may also share common features with the non-perturbative phenomenon of black-hole dominance.

If black-hole states are part of the high-energy spectrum in quadratic gravity, then we might expect some ``non-localizable'' effects to play a key role in the deep UV. Indeed, exponentially suppressed amplitudes cannot be described in a fully local QFT, but tools of non-localizable QFT (in Jaffe's language~\cite{Jaffe:1966an,Jaffe:1967nb}) might be needed~\cite{Keltner:2015xda,Buoninfante:2023dyd,Buoninfante:2024ibt}. This idea is also supported by the fact that spectral densities in local QFT are polynomially bounded, while for black-hole states we would expect an exponentially growing behavior~\cite{Keltner:2015xda}. As a future goal, it would be interesting to rigorously analyse black-hole formation via high-energy scattering and gain a deeper understanding of the bridge between perturbative and non-perturbative sectors and, at the same time, between local and non-localizable dynamics in quantum gravity.

Our discussion of the (super-)Planckian regime of quadratic gravity is still speculative and certainly requires future investigation. However, the main point we wish to emphasize again is that, regardless of what happens at the Planck scale, quadratic gravity describes and predicts new physics beyond GR in the sub-Planckian regime. The question of determining the class of consistent EFTs, which is typically posed in other quantum gravity approaches~\cite{Vafa:2005ui,Basile:2021krr,Eichhorn:2024rkc,Basile:2025zjc},\footnote{It is worth pointing out that quadratic gravity falls into the so-called \textit{swampland}~\cite{Vafa:2005ui,Lust:2023zql}, due to the large values of the quadratic curvature coefficients and the presence of the spin-two ghost. This is good news because we can discriminate between quadratic gravity and other quantum gravity approaches such as string theory. In particular, Starobinsky inflation (driven by $R^2$) cannot be consistently realized in string theory~\cite{Lust:2023zql,Brinkmann:2023eph}. Even if the QFT framework and the strict renormalizability paradigm become inadequate at some high-energy scale, we would certainly be able to claim that string theory cannot be a UV completion of quadratic gravity. In such a case, the \textit{UVland} of quadratic gravity could contain theories and even frameworks yet unknown to us.} does not arise at all if we work in the QFT framework and adopt the strict renormalizability paradigm. As explained in the previous sections, if we remain as conservative as possible we have a unique strictly renormalizable and falsifiable gravitational QFT: quadratic gravity!

\section{An instructive historical note}\label{sec:history}

After the discovery of special relativity and quantum mechanics, it was believed more than once that a new revolution was necessary to make progress in fundamental physics. However, it turned out that no revolution was actually needed, but simply a deeper understanding of what was already available, namely QFT and renormalizability. These two have faced various challenges in the last century, but ultimately overcome them all. Nowadays it seems that the QFT framework and the strict renormalizability paradigm are going through a new phase of challenges related to
understanding gravity at a more fundamental level.

Before concluding, we want to recall some of the past challenges, often forgotten or even unknown to the young generation (as it was for this author), and see if we can learn new lessons that can help us overcome the current ``gravitational'' challenge. We do not aim or pretend to be able to present a complete historical account, but we will simply highlight those aspects that we feel are important for the main message we wish to convey at the end. For more detailed historical accounts of the developments of QFT and SM we refer readers to~\cite{Iliopoulos:2025fhr,Weinberg-history-qft,umezawa-qft,tHooft:2015sdj,Pais1986-PAIIBO,Brown:1993kb,Hoddeson_Brown_Riordan_Dresden_1997,Cao:1997my,galison:1987,kaiser-drawing}.

\subsection*{``Electromagnetic'' challenge}

In the 1930s and most of the 1940s, physicists were struggling to unify special relativity and quantum mechanics into the same framework, especially because of the puzzling UV infinities of QFT. The older generation (especially the founders of quantum mechanics) believed that a new revolution was needed and/or that a more suitable framework needed to be found~\cite{Oppenheimer:1930zz,Dirac:1931kp}.

\medskip

\textbf{Alternative approaches.} In 1938 Werner~Heisenberg considered the existence of a universal minimal length to cutoff the infinities~\cite{Heisenberg-minimal-length}. In 1943 he was the first to propose the ``$S$-matrix theory''~\cite{Heisenberg-S-matrix} (whose concept was introduced by John~Wheeler in 1937~\cite{Wheeler:1937zz}) as an alternative framework that does not involve fields and Lagrangians, but instead deals only with directly-measurable quantities. In 1942 Paul~Dirac considered a generalization of quantum mechanics that included additional intermediate states with negative norms, which could give rise to cancellations in loop diagrams and make the theory finite~\cite{dirac-neg-norm}. Furthermore, Max~Born believed that the laws of nature should be invariant under reciprocity~\cite{Born:1938zve}, i.e. roughly under $x\rightarrow 1/p,$ and was led to propose a non-local theory of electrodynamics through which he wanted to obtain a finite self-energy for the electron~\cite{Born1,Born3,Born:1950iuq}. Hideki~Yukawa followed a similar path~\cite{Yukawa:1950eq,Yukawa:1950er}. 

\medskip

\textbf{QFT and strict renormalizability.} All these alternative attempts failed, in particular none of them could explain the measurement of the Lamb shift~\cite{Lamb:1947zz}, first presented by Willis Lamb at the Shelter Island conference in 1947. During the same conference, Hans Kramers proposed the idea of renormalization that eventually turned out to be the key to making sense of the UV infinities~\cite{Brown:1993kb}. Soon after, Hans Bethe made a non-relativistic computation implementing the renormalization method and could obtain a quantitative result for the Lamb shift that was consistent with the experiment~\cite{Bethe:1947id}. What eventually happened is that QED was shown to be a successful (strictly renormalizable) QFT of electrons, positrons, and photons by Sin-Itiro Tomonaga~\cite{Tomonaga:1946zz,Tomonaga:1948zz}, Julian Schwinger~\cite{Schwinger:1948yk,Schwinger:1948yj,Schwinger:1948iu} and Richard Feynman~\cite{Feynman:1949zx,Feynman:1950ir}. In addition to explaining the Lamb shift, QED was able to make new predictions for the anomalous magnetic moment of the electron~\cite{Schwinger:1948iu}, which was later shown to be in spectacular agreement with experiments. In the same years, Freeman Dyson put the concept of renormalizability by power counting on a more rigorous basis~\cite{Dyson:1949bp,Dyson:1949ha}, and soon after Susumu~Kamefuchi, Shoichi~Sakata and Hiroomi~Umezawa clarified the difference between renormalizable and non-renormalizable QFTs~\cite{Sakata:1952rq}, inspired by Heisenberg's classification of the~interactions~\cite{Heisenberg:1938xla}.

\subsection*{``Electroweak'' and ``strong''  challenges} 

The QED party did not last long, since in the 1950s new challenges arose and plagued QFT for roughly twenty years.  While renormalization theory worked very well for the electromagnetic interaction, it was realized that the same was not true for Fermi's theory of beta decay~\cite{Fermi:1933jpa}. Most of the ingredients of the electroweak theory were already available in the 1960s~\cite{Englert:1964et,Higgs:1964pj,Guralnik:1964eu,Glashow:1961tr,Weinberg:1967tq,Salam:1968rm}, but somehow they were not considered physically relevant because non-Abelian gauge theories were believed to be non-renormalizable. Furthermore, it seemed that the tools of perturbative QFT were not reliable for studying the strong interaction due to the large values of the interaction couplings. At the same time, in 1968 James Bjorken~\cite{Bjorken:1968dy,Bjorken:1969ja}, and subsequently Richard Feynman~\cite{Feynman:1969wa,Feynman:1969ej}, noticed that scaling properties of hadrons in deep inelastic collisions could be well explained by assuming that hadrons behave as objects made of (almost) non-interacting constituents. This behavior, known as Bjorken scaling, seemed to disfavor a possible QFT description of hadrons. Another puzzle was the Sutherland-Veltman paradox~\cite{Sutherland:1967vf,veltman-paradox}, according to which existing theoretical methods were predicting a nearly vanishing rate for the $\pi^0\rightarrow \gamma\gamma$ decay, while experiments were showing the opposite. On top of all these headaches, the discovery of gauge anomalies added further concerns about the internal consistency of the entire renormalization apparatus.

\medskip

\textbf{Alternative approaches.} The situation was really a mess, and it is quite understandable that many physicists began to be seriously skeptical about the whole QFT framework. Lev~Landau noted that known perturbation theories were plagued by Landau poles and became one of the leading figures in believing that a paradigm shift was needed~\cite{Landau:1958jby,Landau:1960iwx}. Alternative approaches that did not involve fields and Lagrangians began to be proposed. The $S$-matrix theory was revived by Geoffrey~Chew in the early 1960s~\cite{Chew:1962mpd} to study strong interactions and its framework was mainly based on dispersion relations~\cite{Chew:1957zz,Mandelstam:1958xc}. This was followed by the dual resonance models and the birth of string theory~\cite{Veneziano:1968yb,Nambu:1969se,Nielsen:1970bc,Susskind:1970qz,Ramond:1971gb,Neveu:1971fz}. 
The framework of current algebra, based on currents instead of fields and on conservation laws instead of Lagrangians, was proposed by Murray~Gell-Mann~\cite{Gell-Mann:1964hhf}. Furthermore, in 1969 Tsung-Dao~Lee and Gian~Carlo~Wick proposed a higher-derivative extension of QED in the hope of obtaining a QFT that could be free of UV divergences~\cite{Lee:1969fy,Lee:1970iw}. Their work caught the attention of various physicists including Sidney~Coleman~\cite{Coleman:1969xz}. In 1971 Garry~Efimov considered a non-local generalization of QED~\cite{Efimov:1971hgl}.

\medskip

\textbf{QFT and strict renormalizability.} During this pessimistic period, characterized by an increasing number of attempts to move away from renormalizable QFTs, new hopes began to arise in the late 1960s. In 1969 John~Bell and Roman~Jackiw~\cite{Bell:1969ts} and Stephen~Adler~\cite{Adler:1969gk} discovered anomalies. In particular, they realized that the negligible value of the rate of the $\pi^0\rightarrow \gamma \gamma$ decay predicted by the current algebra approach~\cite{Sutherland:1967vf,veltman-paradox} was wrong. This had to be corrected by the chiral anomaly contribution, which turned out to be compatible with experiments, thus resolving the Sutherland-Veltman paradox. The amazing aspect of this result is not only that QFT won over current algebra, but also that the chiral anomaly is one-loop exact~\cite{Adler:1969er}, so perturbative QFT was able to provide non-perturbative information on the dynamics of pions. Moreover, the anomalous contribution depends on the number of quarks circulating in the loop, so agreement with experiments requires exactly three generations of quarks.

More successes were yet to come. Martinus~Veltman was perhaps the only one in the 1960s who truly believed that non-Abelian gauge theories could actually describe Nature~\cite{Veltman:1968ki,Reiff:1969pq}. Luckily, he was very stubborn and his intuition was right. Indeed, it was his  student Gerard~'t~Hooft who proved in 1971 that massless~\cite{tHooft:1971akt} and massive non-Abelian gauge theories equipped with the mechanism of spontaneous symmetry breaking~\cite{tHooft:1971qjg} are (strictly) renormalizable. Only after these works did the Glashow-Weinberg-Salam electroweak theory formulated in the 1960s~\cite{Glashow:1961tr,Weinberg:1967tq,Salam:1968rm} began to be taken seriously. Furthermore, in 1972 it was shown that gauge anomalies cancel in the same theory~\cite{Bouchiat:1972iq,Gross:1972pv,Georgi:1972bb}. Thus almost all the pieces of the puzzle seemed to fall~into~place.

Indeed, the puzzle was completed in 1973 by David~Politzer~\cite{Politzer:1973fx} and David~Gross and Frank~Wilczek~\cite{Gross:1973id}, who showed that non-Abelian gauge theories are asymptotically free, which turned out to be the key to a QFT explanation of the Bjorken scaling~\cite{Gross:1973id,Parisi:1973ma,Callan:1973pu}.\footnote{It is worth mentioning that the negative sign of the beta function in non-Abelian gauge theories was already known to Vladimir~Vanyashin and Mikhail~Terentyev in 1965~\cite{Vanyashin:1965ple}, Iosif~Khriplovich in 1969~\cite{Khriplovich:1969aa} and Gerard~'t~Hooft in 1971~\cite{Veltman:2000xp,shifman2024historicalcuriosityasymptoticfreedom}. However, at that time the physical implications of the result were not understood.} This established QCD~\cite{Fritzsch:1972jv,Fritzsch:1973pi} as the (strictly renormalizable) QFT of strong interaction. In the same year 1973 the observation of neutral currents marked the first experimental verification of the electroweak theory~\cite{Hasert:1973cr}. Since then, many other experiments have confirmed~the~SM~predictions~\cite{Iliopoulos:2025fhr}.

\subsection*{``Gravitational'' challenge}

After the SM of particle physics was formulated, the next obvious question to ask was whether quantum aspects of the gravitational interaction could be described with the same QFT tools. It had already been suspected that GR was perturbatively non-renormalizable since the middle of the last century~\cite{Gupta:1952zz,Utiyama:1962sn}, but it was not until 1974 that Gerard~'t Hooft and Martinus~Veltman~\cite{tHooft:1974toh} performed the first complete one-loop calculation of the UV divergences. They showed that pure Einstein's gravity is one-loop renormalizable, but when the coupling with matter is switched on one-loop renormalizability is lost. Later, in 1985 Marc~Goroff and Augusto~Sagnotti~\cite{Goroff:1985sz} calculated the non-renormalizable two-loop divergence in pure gravity. At this point, knowing that the SM Lagrangian is strictly renormalizable, one should still ask whether something similar can be obtained for gravity, analogously to what happened with the transition from Fermi's theory to electroweak completion. Around 1974, Stanley~Deser~\cite{Deser:1974zz} and Steven~Weinberg~\cite{Weinberg:1974tw} were the first to propose quadratic gravity as a strictly renormalizable QFT of gravity. They used only power counting arguments, but the complete proof of renormalizability was performed by Kellog~Stelle in 1976~\cite{Stelle:1976gc}. Quadratic gravity was further studied in the following years~\cite{Tomboulis:1977jk,Julve:1978xn,Fradkin:1981iu,Avramidi:1985ki,Salam:1978fd,Tomboulis:1980bs,Tomboulis:1983sw,Adler:1982ri,Strominger:1982wp,Boulware:1983yj,Boulware:1983vw,Hawking:1985gh,Antoniadis:1986tu,Johnston:1987ue}, but it was soon abandoned because most physicists of that time were afraid of ghosts and, moreover, the potential phenomenological implications of the theory had not yet been appreciated. In fact, quadratic gravity slowly started to be forgotten, while in those years alternative approaches, paradigms and  frameworks for quantum gravity were proposed.

\medskip

\textbf{Alternative approaches}. From 1974 and especially in the 1980s (super)string theory had a revival and with it the $S$-matrix bootstrap again~\cite{Scherk:1974ca,Yoneya:1974jg,Green:1981yb,Green:1984sg,Gross:1984dd,Candelas:1985en}. The idea of supergravity started to become popular in 1976~\cite{Freedman:1976xh}. In the same years Steven~Weinberg also tried to depart from the standard framework of perturbative QFT and proposed what is known as asymptotically safe quantum gravity~\cite{Weinberg:1976xy,Weinberg:1980gg}. According to this approach, the EFT Lagrangian of GR, containing infinitely many (perturbatively) non-renormalizable operators, could eventually be non-perturbatively renormalizable, if there existed an interacting UV fixed point with a finite-dimensional critical surface~\cite{Donoghue:2019clr,Bonanno:2020bil}. It was during this period that the Wilsonian view of renormalizability began to spread in the high-energy physics community, and with it the idea that all QFTs, including the strictly renormalizable ones, could simply be EFTs~\cite{Weinberg:1978kz,Weinberg:1996kw,Weinberg:2009bg,Rocci:2024vrq}. In the second half of the 1980s, approaches aiming at a fully non-perturbative (discrete) quantization of spacetime were proposed, such as loop quantum gravity~\cite{Ashtekar:1986yd,Ashtekar:1987gu,Rovelli:1989za} and causal set theory~\cite{Bombelli:1987aa}. In 1987 the idea of non-local field theories was again considered, this time applied to gravity~\cite{Krasnikov:1987yj,Kuzmin:1989sp}. Nowadays all these approaches, in addition to new ones, are still  actively studied~\cite{Buoninfante:2024yth}. 

\medskip

\textbf{QFT and strict renormalizability again.} The community of QFT in curved spacetime knew already in the early 1980s that quadratic terms in the curvature could have interesting phenomenological implications for the early universe cosmology, especially thanks to Alexei~Starobinsky's inflationary model~\cite{Starobinsky:1980te,Mukhanov:1981xt,Starobinsky:1981vz,Starobinsky:1983zz,Kofman:1985aw,Vilenkin:1985md}. However, the history of those years evolved in such a way that the quantum gravity community did not try to take advantage of it, probably because concrete observational data were still lacking. In fact, it was only after the 2012 WMAP data~\cite{WMAP:2012nax} and especially after the 2013 Planck data~\cite{Planck:2013jfk} that Starobinsky inflation began to be taken very seriously. Shortly thereafter, quadratic gravity -- namely the old lore of quantizing gravity in the framework of perturbative QFT consistently with the strict renormalizability criterion -- began to experience a renaissance. This started with the works of Alberto~Salvio~\cite{Salvio:2014soa,Salvio:2016vxi,Salvio:2019ewf,Salvio:2018kwh,Salvio:2018crh,Salvio:2020axm}, Bob Holdom~\cite{Holdom:2015kbf,Holdom:2021hlo,Holdom:2021oii}, Damiano~Anselmi and Marco~Piva~\cite{Anselmi:2017ygm,Anselmi:2018ibi,Anselmi:2018tmf,Anselmi:2018bra}, John~Donoghue and Gabriel~Menezes~\cite{Donoghue:2018izj,Donoghue:2019ecz,Donoghue:2019fcb,Donoghue:2021cza,Donoghue:2021meq}. As discussed in this paper, new insights into the true nature of the spin-two ghost are now available, and in addition to the successful description of the CMB anisotropies, quadratic gravity provides cosmological predictions that can be confirmed or falsified in the near future.

\subsection*{Take-home message}

What lessons can we learn from this historical account? We can certainly say that in the past century, whenever the perturbative QFT framework has been challenged, it has ultimately proven triumphant, and with it the strict renormalizability paradigm. This does not mean that all the ideas behind the alternative approaches were useless. For example, Dirac's tools for dealing with negative norms have provided useful methods for formulating gauge theories. Moreover, results from current algebra and $S$-matrix bootstrap have inspired the later QFT successes.

A striking fact that we must observe is that most of the approaches proposed as alternatives to the perturbative QFT framework during the three different phases of challenges  share similar features and many of them have seen a revival. For example, Heisenberg's 1938 idea of the existence of a universal minimal length has seen a significant revival in the context of quantum gravity, especially in the case of discrete approaches~\cite{Loll:1998aj}. The $S$-matrix bootstrap has continued to be reconsidered in all three phases, despite its proven impotence against QFT, and is perhaps having its greatest revival today~\cite{Kruczenski:2022lot}. After failing to describe the strong interaction, string theory quickly became one of the most popular approaches to quantum gravity. Another recurring theme throughout the three phases has been that of non-local field theories, and indeed non-local quantum gravity is still being actively studied today~\cite{BasiBeneito:2022wux}.

Is history repeating itself once again? Only time will tell, but for now we can objectively state that history and current observations stubbornly point towards a strictly renormalizable QFT description of gravity. Furthermore, quadratic gravity makes falsifiable cosmological predictions. The latter, if verified observationally, could represent a new confirmation of the strict renormalizability paradigm. In this case we might actually start to think that the old property of ``being strictly renormalizable'' (contrary to the modern Wilsonian view) is not just a matter of luck~\cite{Butterfield:2014rxa,Butterfield:2014oja}, but may have a much deeper meaning. Even if  the strict renormalizability paradigm does not turn out to be the final step in the evolutionary process of fundamental theoretical physics, history and Nature are telling us that we cannot proceed without~it.

In a recent meeting on quantum gravity~\cite{Buoninfante:2024yth}, John~Donoghue raised an interesting  \href{https://youtu.be/3_L1MFf1BrU?t=3310}{question}: ``What is success [for quadratic gravity]?'' Our answer would be that the uniqueness of quadratic gravity as the most conservative renormalizable QFT of gravity, its successful explanation of the CMB anisotropies, and its sharp predictions constitute an unprecedented success for the QFT framework applied to gravity.
It is precisely this success that should give us the right motivation to delve deeper into the theory and try to find definitive answers to the open questions about the spin-two ghost, microcausality and the (super-)Planckian regime. We do not know where this journey will take us, but it is definitely worth getting on board and continuing the adventure.


\subsection*{Acknowledgements}

I would like to thank Damiano Anselmi, Jeremy Butterfield, Francesco Di Filippo, John Donoghue, Antonio Ferreiro, Bob Holdom, Gerard 't Hooft, Sravan Kumar, Roberto Percacci, Alberto Salvio, and Fabio Scardigli for useful comments and suggestions on the manuscript. 
I am grateful to Sravan Kumar for spotting typos and Jeremy Butterfield for valuable linguistic tips.
I would also like to thank Ivano~Basile, Benjamin~Knorr, and Alessia~Platania for enlightening discussions and constructive criticisms.
I acknowledge financial support from the European Union’s Horizon 2020 research and innovation programme under the Marie Skłodowska-Curie Actions (grant agreement ID:~101106345-NLQG).
	


\bibliographystyle{utphys}
\bibliography{References}

\providecommand{\href}[2]{#2}\begingroup\raggedright\begin{thebibliography}{100}

\bibitem{Iliopoulos:2025fhr}
J.~Iliopoulos, ``{From Many Models to ONE THEORY},'' \href{http://arxiv.org/abs/2501.10233}{{\ttfamily arXiv:2501.10233 [physics.hist-ph]}}.

\bibitem{Dyson:1949bp}
F.~J. Dyson, ``{The Radiation theories of Tomonaga, Schwinger, and Feynman},'' \href{http://dx.doi.org/10.1103/PhysRev.75.486}{{\em Phys. Rev.} {\bfseries 75} (1949) 486--502}.

\bibitem{Dyson:1949ha}
F.~J. Dyson, ``{The S matrix in quantum electrodynamics},'' \href{http://dx.doi.org/10.1103/PhysRev.75.1736}{{\em Phys. Rev.} {\bfseries 75} (1949) 1736--1755}.

\bibitem{Sakata:1952rq}
S.~Sakata, H.~Umezawa, and S.~Kamefuchi, ``{On the structure of the interaction of the elementary particles. I: The renormalizability of the interaction},'' \href{http://dx.doi.org/10.1143/PTP.7.377}{{\em Prog. Theor. Phys.} {\bfseries 7} (1952) 377--390}.

\bibitem{Basile:2024oms}
I.~Basile, L.~Buoninfante, F.~Di~Filippo, B.~Knorr, A.~Platania, and A.~Tokareva, {\em {Lectures in Quantum Gravity}}.
\newblock 12, 2024.
\newblock \href{http://arxiv.org/abs/2412.08690}{{\ttfamily arXiv:2412.08690 [hep-th]}}.

\bibitem{Anselmi:1994ry}
D.~Anselmi, ``{Removal of divergences with the Batalin-Vilkovisky formalism},'' \href{http://dx.doi.org/10.1088/0264-9381/11/9/005}{{\em Class. Quant. Grav.} {\bfseries 11} (1994) 2181--2204}.

\bibitem{Gomis:1995jp}
J.~Gomis and S.~Weinberg, ``{Are nonrenormalizable gauge theories renormalizable?},'' \href{http://dx.doi.org/10.1016/0550-3213(96)00132-0}{{\em Nucl. Phys. B} {\bfseries 469} (1996) 473--487}, \href{http://arxiv.org/abs/hep-th/9510087}{{\ttfamily arXiv:hep-th/9510087}}.

\bibitem{Tomonaga:1946zz}
S.~Tomonaga, ``{On a relativistically invariant formulation of the quantum theory of wave fields},'' \href{http://dx.doi.org/10.1143/PTP.1.27}{{\em Prog. Theor. Phys.} {\bfseries 1} (1946) 27--42}.

\bibitem{Tomonaga:1948zz}
S.-I. Tomonaga, ``{On Infinite Field Reactions in Quantum Field Theory},'' \href{http://dx.doi.org/10.1103/PhysRev.74.224}{{\em Phys. Rev.} {\bfseries 74} (1948) 224--225}.

\bibitem{Schwinger:1948yk}
J.~S. Schwinger, ``{Quantum electrodynamics. I A covariant formulation},'' \href{http://dx.doi.org/10.1103/PhysRev.74.1439}{{\em Phys. Rev.} {\bfseries 74} (1948) 1439}.

\bibitem{Schwinger:1948yj}
J.~S. Schwinger, ``{Quantum electrodynamics. 2. Vacuum polarization and selfenergy},'' \href{http://dx.doi.org/10.1103/PhysRev.75.651}{{\em Phys. Rev.} {\bfseries 75} (1948) 651}.

\bibitem{Schwinger:1948iu}
J.~S. Schwinger, ``{On Quantum electrodynamics and the magnetic moment of the electron},'' \href{http://dx.doi.org/10.1103/PhysRev.73.416}{{\em Phys. Rev.} {\bfseries 73} (1948) 416--417}.

\bibitem{Feynman:1949zx}
R.~P. Feynman, ``{Space - time approach to quantum electrodynamics},'' \href{http://dx.doi.org/10.1103/PhysRev.76.769}{{\em Phys. Rev.} {\bfseries 76} (1949) 769--789}.

\bibitem{Feynman:1950ir}
R.~P. Feynman, ``{Mathematical formulation of the quantum theory of electromagnetic interaction},'' \href{http://dx.doi.org/10.1103/PhysRev.80.440}{{\em Phys. Rev.} {\bfseries 80} (1950) 440--457}.

\bibitem{Fermi:1933jpa}
E.~Fermi, ``{Tentativo di una teoria dell'emissione dei raggi beta},'' {\em Ric. Sci.} {\bfseries 4} (1933) 491--495.

\bibitem{Englert:1964et}
F.~Englert and R.~Brout, ``{Broken Symmetry and the Mass of Gauge Vector Mesons},'' \href{http://dx.doi.org/10.1103/PhysRevLett.13.321}{{\em Phys. Rev. Lett.} {\bfseries 13} (1964) 321--323}.

\bibitem{Higgs:1964pj}
P.~W. Higgs, ``{Broken Symmetries and the Masses of Gauge Bosons},'' \href{http://dx.doi.org/10.1103/PhysRevLett.13.508}{{\em Phys. Rev. Lett.} {\bfseries 13} (1964) 508--509}.

\bibitem{Guralnik:1964eu}
G.~S. Guralnik, C.~R. Hagen, and T.~W.~B. Kibble, ``{Global Conservation Laws and Massless Particles},'' \href{http://dx.doi.org/10.1103/PhysRevLett.13.585}{{\em Phys. Rev. Lett.} {\bfseries 13} (1964) 585--587}.

\bibitem{Glashow:1961tr}
S.~L. Glashow, ``{Partial Symmetries of Weak Interactions},'' \href{http://dx.doi.org/10.1016/0029-5582(61)90469-2}{{\em Nucl. Phys.} {\bfseries 22} (1961) 579--588}.

\bibitem{Weinberg:1967tq}
S.~Weinberg, ``{A Model of Leptons},'' \href{http://dx.doi.org/10.1103/PhysRevLett.19.1264}{{\em Phys. Rev. Lett.} {\bfseries 19} (1967) 1264--1266}.

\bibitem{Salam:1968rm}
A.~Salam, ``{Weak and Electromagnetic Interactions},'' \href{http://dx.doi.org/10.1142/9789812795915_0034}{{\em Conf. Proc. C} {\bfseries 680519} (1968) 367--377}.

\bibitem{tHooft:1971akt}
G.~'t~Hooft, ``{Renormalization of Massless Yang-Mills Fields},'' \href{http://dx.doi.org/10.1016/0550-3213(71)90395-6}{{\em Nucl. Phys. B} {\bfseries 33} (1971) 173--199}.

\bibitem{tHooft:1971qjg}
G.~'t~Hooft, ``{Renormalizable Lagrangians for Massive Yang-Mills Fields},'' \href{http://dx.doi.org/10.1016/0550-3213(71)90139-8}{{\em Nucl. Phys. B} {\bfseries 35} (1971) 167--188}.

\bibitem{Fritzsch:1972jv}
H.~Fritzsch and M.~Gell-Mann, ``{Current algebra: Quarks and what else?},'' {\em eConf} {\bfseries C720906V2} (1972) 135--165, \href{http://arxiv.org/abs/hep-ph/0208010}{{\ttfamily arXiv:hep-ph/0208010}}.

\bibitem{Fritzsch:1973pi}
H.~Fritzsch, M.~Gell-Mann, and H.~Leutwyler, ``{Advantages of the Color Octet Gluon Picture},'' \href{http://dx.doi.org/10.1016/0370-2693(73)90625-4}{{\em Phys. Lett. B} {\bfseries 47} (1973) 365--368}.

\bibitem{Politzer:1973fx}
H.~D. Politzer, ``{Reliable Perturbative Results for Strong Interactions?},'' \href{http://dx.doi.org/10.1103/PhysRevLett.30.1346}{{\em Phys. Rev. Lett.} {\bfseries 30} (1973) 1346--1349}.

\bibitem{Gross:1973id}
D.~J. Gross and F.~Wilczek, ``{Ultraviolet Behavior of Nonabelian Gauge Theories},'' \href{http://dx.doi.org/10.1103/PhysRevLett.30.1343}{{\em Phys. Rev. Lett.} {\bfseries 30} (1973) 1343--1346}.

\bibitem{tHooft:1974toh}
G.~'t~Hooft and M.~J.~G. Veltman, ``{One loop divergencies in the theory of gravitation},''
{\em Ann. Inst. H. Poincare Phys. Theor.} {\bfseries A20} (1974) 69--94.

\bibitem{Goroff:1985sz}
M.~H. Goroff and A.~Sagnotti, ``{QUANTUM GRAVITY AT TWO LOOPS},''
\href{http://dx.doi.org/10.1016/0370-2693(85)91470-4}{{\em Phys. Lett.} {\bfseries 160B} (1985) 81--86}.

\bibitem{Donoghue:1994dn}
J.~F. Donoghue, ``{General relativity as an effective field theory: The leading quantum corrections},'' \href{http://dx.doi.org/10.1103/PhysRevD.50.3874}{{\em Phys. Rev.} {\bfseries D50} (1994) 3874--3888},
\href{http://arxiv.org/abs/gr-qc/9405057}{{\ttfamily arXiv:gr-qc/9405057 [gr-qc]}}.

\bibitem{Han:2004wt}
T.~Han and S.~Willenbrock, ``{Scale of quantum gravity},'' \href{http://dx.doi.org/10.1016/j.physletb.2005.04.040}{{\em Phys. Lett. B} {\bfseries 616} (2005) 215--220}, \href{http://arxiv.org/abs/hep-ph/0404182}{{\ttfamily arXiv:hep-ph/0404182}}.

\bibitem{Dvali:2007hz}
G.~Dvali, ``{Black Holes and Large N Species Solution to the Hierarchy Problem},'' \href{http://dx.doi.org/10.1002/prop.201000009}{{\em Fortsch. Phys.} {\bfseries 58} (2010) 528--536}, \href{http://arxiv.org/abs/0706.2050}{{\ttfamily arXiv:0706.2050 [hep-th]}}.

\bibitem{Dvali:2007wp}
G.~Dvali and M.~Redi, ``{Black Hole Bound on the Number of Species and Quantum Gravity at LHC},'' \href{http://dx.doi.org/10.1103/PhysRevD.77.045027}{{\em Phys. Rev. D} {\bfseries 77} (2008) 045027}, \href{http://arxiv.org/abs/0710.4344}{{\ttfamily arXiv:0710.4344 [hep-th]}}.

\bibitem{Rocci:2024vrq}
A.~Rocci and T.~Van~Riet, ``{The quantum theory of gravitation, effective field theories, and strings: yesterday and today},'' \href{http://dx.doi.org/10.1140/epjh/s13129-024-00069-4}{{\em Eur. Phys. J. H} {\bfseries 49} no.~1, (2024) 7}, \href{http://arxiv.org/abs/2403.14008}{{\ttfamily arXiv:2403.14008 [physics.hist-ph]}}.

\bibitem{WALLACE202231}
D.~Wallace, ``Quantum gravity at low energies,'' \href{http://dx.doi.org/https://doi.org/10.1016/j.shpsa.2022.04.003}{{\em Studies in History and Philosophy of Science} {\bfseries 94} (2022) 31--46}, \href{http://arxiv.org/abs/2112.12235}{{\ttfamily arXiv:2112.12235 [gr-qc]}}.

\bibitem{Bonanno:2020bil}
A.~Bonanno, A.~Eichhorn, H.~Gies, J.~M. Pawlowski, R.~Percacci, M.~Reuter, F.~Saueressig, and G.~P. Vacca, ``{Critical reflections on asymptotically safe gravity},'' \href{http://dx.doi.org/10.3389/fphy.2020.00269}{{\em Front. in Phys.} {\bfseries 8} (2020) 269}, \href{http://arxiv.org/abs/2004.06810}{{\ttfamily arXiv:2004.06810 [gr-qc]}}.

\bibitem{Loll:2019rdj}
R.~Loll, ``{Quantum Gravity from Causal Dynamical Triangulations: A Review},'' \href{http://dx.doi.org/10.1088/1361-6382/ab57c7}{{\em Class. Quant. Grav.} {\bfseries 37} no.~1, (2020) 013002}, \href{http://arxiv.org/abs/1905.08669}{{\ttfamily arXiv:1905.08669 [hep-th]}}.

\bibitem{Surya:2019ndm}
S.~Surya, ``{The causal set approach to quantum gravity},'' \href{http://dx.doi.org/10.1007/s41114-019-0023-1}{{\em Living Rev. Rel.} {\bfseries 22} no.~1, (2019) 5}, \href{http://arxiv.org/abs/1903.11544}{{\ttfamily arXiv:1903.11544 [gr-qc]}}.

\bibitem{Stelle:1976gc}
K.~S. Stelle, ``{Renormalization of Higher Derivative Quantum Gravity},'' \href{http://dx.doi.org/10.1103/PhysRevD.16.953}{{\em Phys. Rev. D} {\bfseries 16} (1977) 953--969}.

\bibitem{Tomboulis:1977jk}
E.~Tomboulis, ``{1/N Expansion and Renormalization in Quantum Gravity},'' \href{http://dx.doi.org/10.1016/0370-2693(77)90678-5}{{\em Phys. Lett. B} {\bfseries 70} (1977) 361--364}.

\bibitem{Julve:1978xn}
J.~Julve and M.~Tonin, ``{Quantum Gravity with Higher Derivative Terms},'' \href{http://dx.doi.org/10.1007/BF02748637}{{\em Nuovo Cim. B} {\bfseries 46} (1978) 137--152}.

\bibitem{Fradkin:1981iu}
E.~S. Fradkin and A.~A. Tseytlin, ``{Renormalizable asymptotically free quantum theory of gravity},'' \href{http://dx.doi.org/10.1016/0550-3213(82)90444-8}{{\em Nucl. Phys. B} {\bfseries 201} (1982) 469--491}.

\bibitem{Avramidi:1985ki}
I.~G. Avramidi and A.~O. Barvinsky, ``{ASYMPTOTIC FREEDOM IN HIGHER DERIVATIVE QUANTUM GRAVITY},'' \href{http://dx.doi.org/10.1016/0370-2693(85)90248-5}{{\em Phys. Lett. B} {\bfseries 159} (1985) 269--274}.

\bibitem{Salam:1978fd}
A.~Salam and J.~A. Strathdee, ``{Remarks on High-energy Stability and Renormalizability of Gravity Theory},'' \href{http://dx.doi.org/10.1103/PhysRevD.18.4480}{{\em Phys. Rev. D} {\bfseries 18} (1978) 4480}.

\bibitem{Tomboulis:1980bs}
E.~Tomboulis, ``{Renormalizability and Asymptotic Freedom in Quantum Gravity},'' \href{http://dx.doi.org/10.1016/0370-2693(80)90550-X}{{\em Phys. Lett. B} {\bfseries 97} (1980) 77--80}.

\bibitem{Tomboulis:1983sw}
E.~T. Tomboulis, ``{Unitarity in Higher Derivative Quantum Gravity},'' \href{http://dx.doi.org/10.1103/PhysRevLett.52.1173}{{\em Phys. Rev. Lett.} {\bfseries 52} (1984) 1173}.

\bibitem{Salvio:2016vxi}
A.~Salvio, ``{Solving the Standard Model Problems in Softened Gravity},'' \href{http://dx.doi.org/10.1103/PhysRevD.94.096007}{{\em Phys. Rev. D} {\bfseries 94} no.~9, (2016) 096007}, \href{http://arxiv.org/abs/1608.01194}{{\ttfamily arXiv:1608.01194 [hep-ph]}}.

\bibitem{Salvio:2019ewf}
A.~Salvio, ``{Metastability in Quadratic Gravity},'' \href{http://dx.doi.org/10.1103/PhysRevD.99.103507}{{\em Phys. Rev. D} {\bfseries 99} no.~10, (2019) 103507}, \href{http://arxiv.org/abs/1902.09557}{{\ttfamily arXiv:1902.09557 [gr-qc]}}.

\bibitem{Salvio:2018kwh}
A.~Salvio, A.~Strumia, and H.~Veerm\"ae, ``{New infra-red enhancements in 4-derivative gravity},'' \href{http://dx.doi.org/10.1140/epjc/s10052-018-6311-1}{{\em Eur. Phys. J. C} {\bfseries 78} no.~10, (2018) 842}, \href{http://arxiv.org/abs/1808.07883}{{\ttfamily arXiv:1808.07883 [hep-th]}}.

\bibitem{Salvio:2018crh}
A.~Salvio, ``{Quadratic Gravity},'' \href{http://dx.doi.org/10.3389/fphy.2018.00077}{{\em Front. in Phys.} {\bfseries 6} (2018) 77}, \href{http://arxiv.org/abs/1804.09944}{{\ttfamily arXiv:1804.09944 [hep-th]}}.

\bibitem{Salvio:2020axm}
A.~Salvio, ``{Dimensional Transmutation in Gravity and Cosmology},'' \href{http://dx.doi.org/10.1142/S0217751X21300064}{{\em Int. J. Mod. Phys. A} {\bfseries 36} no.~08n09, (2021) 2130006}, \href{http://arxiv.org/abs/2012.11608}{{\ttfamily arXiv:2012.11608 [hep-th]}}.

\bibitem{Salvio:2024joi}
A.~Salvio, ``{A non-perturbative and background-independent formulation of quadratic gravity},'' \href{http://dx.doi.org/10.1088/1475-7516/2024/07/092}{{\em JCAP} {\bfseries 07} (2024) 092}, \href{http://arxiv.org/abs/2404.08034}{{\ttfamily arXiv:2404.08034 [hep-th]}}.

\bibitem{Holdom:2015kbf}
B.~Holdom and J.~Ren, ``{QCD analogy for quantum gravity},'' \href{http://dx.doi.org/10.1103/PhysRevD.93.124030}{{\em Phys. Rev. D} {\bfseries 93} no.~12, (2016) 124030}, \href{http://arxiv.org/abs/1512.05305}{{\ttfamily arXiv:1512.05305 [hep-th]}}.

\bibitem{Holdom:2021hlo}
B.~Holdom, ``{Ultra-Planckian scattering from a QFT for gravity},'' \href{http://dx.doi.org/10.1103/PhysRevD.105.046008}{{\em Phys. Rev. D} {\bfseries 105} no.~4, (2022) 046008}, \href{http://arxiv.org/abs/2107.01727}{{\ttfamily arXiv:2107.01727 [hep-th]}}.

\bibitem{Holdom:2021oii}
B.~Holdom, ``{Photon-photon scattering from a UV-complete gravity QFT},'' \href{http://dx.doi.org/10.1007/JHEP04(2022)133}{{\em JHEP} {\bfseries 04} (2022) 133}, \href{http://arxiv.org/abs/2110.02246}{{\ttfamily arXiv:2110.02246 [hep-ph]}}.

\bibitem{Holdom:2023usn}
B.~Holdom, ``{Running couplings and unitarity in a 4-derivative scalar field theory},'' \href{http://dx.doi.org/10.1016/j.physletb.2023.138023}{{\em Phys. Lett. B} {\bfseries 843} (2023) 138023}, \href{http://arxiv.org/abs/2303.06723}{{\ttfamily arXiv:2303.06723 [hep-th]}}.

\bibitem{Holdom:2024cfq}
B.~Holdom, ``{UV-complete 4-derivative scalar field theory},'' \href{http://dx.doi.org/10.1016/j.nuclphysb.2024.116472}{{\em Nucl. Phys. B} {\bfseries 1000} (2024) 116472}, \href{http://arxiv.org/abs/2402.09223}{{\ttfamily arXiv:2402.09223 [hep-th]}}.

\bibitem{Anselmi:2017ygm}
D.~Anselmi, ``{On the quantum field theory of the gravitational interactions},'' \href{http://dx.doi.org/10.1007/JHEP06(2017)086}{{\em JHEP} {\bfseries 06} (2017) 086},
\href{http://arxiv.org/abs/1704.07728}{{\ttfamily arXiv:1704.07728 [hep-th]}}.

\bibitem{Anselmi:2018ibi}
D.~Anselmi and M.~Piva, ``{The Ultraviolet Behavior of Quantum Gravity},'' \href{http://dx.doi.org/10.1007/JHEP05(2018)027}{{\em JHEP} {\bfseries 05} (2018) 027},
\href{http://arxiv.org/abs/1803.07777}{{\ttfamily arXiv:1803.07777 [hep-th]}}.

\bibitem{Anselmi:2018tmf}
D.~Anselmi and M.~Piva, ``{Quantum Gravity, Fakeons And Microcausality},'' \href{http://dx.doi.org/10.1007/JHEP11(2018)021}{{\em JHEP} {\bfseries 11} (2018) 021},
\href{http://arxiv.org/abs/1806.03605}{{\ttfamily arXiv:1806.03605 [hep-th]}}.

\bibitem{Anselmi:2018bra}
D.~Anselmi, ``{Fakeons, Microcausality And The Classical Limit Of Quantum Gravity},'' \href{http://dx.doi.org/10.1088/1361-6382/ab04c8}{{\em Class. Quant. Grav.} {\bfseries 36} (2019) 065010},
\href{http://arxiv.org/abs/1809.05037}{{\ttfamily arXiv:1809.05037 [hep-th]}}.

\bibitem{Piva:2023bcf}
M.~Piva, ``{Higher-derivative quantum gravity with purely virtual particles: renormalizability and unitarity},'' \href{http://dx.doi.org/10.1140/epjp/s13360-023-04486-0}{{\em Eur. Phys. J. Plus} {\bfseries 138} no.~10, (2023) 876}, \href{http://arxiv.org/abs/2305.12549}{{\ttfamily arXiv:2305.12549 [hep-th]}}.

\bibitem{Donoghue:2018izj}
J.~F. Donoghue and G.~Menezes, ``{Gauge Assisted Quadratic Gravity: A Framework for UV Complete Quantum Gravity},'' \href{http://dx.doi.org/10.1103/PhysRevD.97.126005}{{\em Phys. Rev. D} {\bfseries 97} no.~12, (2018) 126005}, \href{http://arxiv.org/abs/1804.04980}{{\ttfamily arXiv:1804.04980 [hep-th]}}.

\bibitem{Donoghue:2019ecz}
J.~F. Donoghue and G.~Menezes, ``{Arrow of Causality and Quantum Gravity},'' \href{http://dx.doi.org/10.1103/PhysRevLett.123.171601}{{\em Phys. Rev. Lett.} {\bfseries 123} no.~17, (2019) 171601}, \href{http://arxiv.org/abs/1908.04170}{{\ttfamily arXiv:1908.04170 [hep-th]}}.

\bibitem{Donoghue:2019fcb}
J.~F. Donoghue and G.~Menezes, ``{Unitarity, stability and loops of unstable ghosts},'' \href{http://dx.doi.org/10.1103/PhysRevD.100.105006}{{\em Phys. Rev. D} {\bfseries 100} no.~10, (2019) 105006}, \href{http://arxiv.org/abs/1908.02416}{{\ttfamily arXiv:1908.02416 [hep-th]}}.

\bibitem{Donoghue:2021cza}
J.~F. Donoghue and G.~Menezes, ``{On quadratic gravity},'' \href{http://dx.doi.org/10.1393/ncc/i2022-22026-7}{{\em Nuovo Cim. C} {\bfseries 45} no.~2, (2022) 26}, \href{http://arxiv.org/abs/2112.01974}{{\ttfamily arXiv:2112.01974 [hep-th]}}.

\bibitem{Donoghue:2021meq}
J.~F. Donoghue and G.~Menezes, ``{Causality and gravity},'' \href{http://dx.doi.org/10.1007/JHEP11(2021)010}{{\em JHEP} {\bfseries 11} (2021) 010}, \href{http://arxiv.org/abs/2106.05912}{{\ttfamily arXiv:2106.05912 [hep-th]}}.

\bibitem{Buoninfante:2023ryt}
L.~Buoninfante, ``{Massless and partially massless limits in Quadratic Gravity},'' \href{http://dx.doi.org/10.1007/JHEP12(2023)111}{{\em JHEP} {\bfseries 12} (2023) 111}, \href{http://arxiv.org/abs/2308.11324}{{\ttfamily arXiv:2308.11324 [hep-th]}}.

\bibitem{Anber:2011ut}
M.~M. Anber and J.~F. Donoghue, ``{On the running of the gravitational constant},'' \href{http://dx.doi.org/10.1103/PhysRevD.85.104016}{{\em Phys. Rev. D} {\bfseries 85} (2012) 104016}, \href{http://arxiv.org/abs/1111.2875}{{\ttfamily arXiv:1111.2875 [hep-th]}}.

\bibitem{Donoghue:2024uay}
J.~F. Donoghue, ``{Do $\Lambda_{CC}$ and $G$ run?},'' in {\em {64th Cracow School of Theoretical Physics From the UltraViolet to the InfraRed}: {A panorama of modern gravitational physics}}.
\newblock 12, 2024.
\newblock \href{http://arxiv.org/abs/2412.08773}{{\ttfamily arXiv:2412.08773 [hep-th]}}.

\bibitem{Buccio:2024hys}
D.~Buccio, J.~F. Donoghue, G.~Menezes, and R.~Percacci, ``{Physical Running of Couplings in Quadratic Gravity},'' \href{http://dx.doi.org/10.1103/PhysRevLett.133.021604}{{\em Phys. Rev. Lett.} {\bfseries 133} no.~2, (2024) 021604}, \href{http://arxiv.org/abs/2403.02397}{{\ttfamily arXiv:2403.02397 [hep-th]}}.

\bibitem{Weinberg:1978kz}
S.~Weinberg, ``{Phenomenological Lagrangians},'' \href{http://dx.doi.org/10.1016/0378-4371(79)90223-1}{{\em Physica A} {\bfseries 96} no.~1-2, (1979) 327--340}.

\bibitem{Weinberg:1996kw}
S.~Weinberg, ``{What is quantum field theory, and what did we think it is?},'' in {\em {Conference on Historical Examination and Philosophical Reflections on the Foundations of Quantum Field Theory}}, pp.~241--251.
\newblock 3, 1996.
\newblock \href{http://arxiv.org/abs/hep-th/9702027}{{\ttfamily arXiv:hep-th/9702027}}.

\bibitem{Weinberg:2009bg}
S.~Weinberg, ``{Effective Field Theory, Past and Future},'' \href{http://dx.doi.org/10.22323/1.086.0001}{{\em PoS} {\bfseries CD09} (2009) 001}, \href{http://arxiv.org/abs/0908.1964}{{\ttfamily arXiv:0908.1964 [hep-th]}}.

\bibitem{Salvio:2014soa}
A.~Salvio and A.~Strumia, ``{Agravity},'' \href{http://dx.doi.org/10.1007/JHEP06(2014)080}{{\em JHEP} {\bfseries 06} (2014) 080}, \href{http://arxiv.org/abs/1403.4226}{{\ttfamily arXiv:1403.4226 [hep-ph]}}.

\bibitem{Mannheim:2011ds}
P.~D. Mannheim, ``{Making the Case for Conformal Gravity},'' \href{http://dx.doi.org/10.1007/s10701-011-9608-6}{{\em Found. Phys.} {\bfseries 42} (2012) 388--420}, \href{http://arxiv.org/abs/1101.2186}{{\ttfamily arXiv:1101.2186 [hep-th]}}.

\bibitem{tHooft:2015tlq}
G.~'t~Hooft, ``{Singularities, horizons, firewalls, and local conformal symmetry},'' \href{http://dx.doi.org/10.1007/978-3-319-94256-8_1}{{\em Springer Proc. Phys.} {\bfseries 208} (2018) 1--12}, \href{http://arxiv.org/abs/1511.04427}{{\ttfamily arXiv:1511.04427 [gr-qc]}}.

\bibitem{Jizba:2014taa}
P.~Jizba, H.~Kleinert, and F.~Scardigli, ``{Inflationary cosmology from quantum Conformal Gravity},'' \href{http://dx.doi.org/10.1140/epjc/s10052-015-3441-6}{{\em Eur. Phys. J. C} {\bfseries 75} no.~6, (2015) 245}, \href{http://arxiv.org/abs/1410.8062}{{\ttfamily arXiv:1410.8062 [hep-th]}}.

\bibitem{Fradkin:1983tg}
E.~S. Fradkin and A.~A. Tseytlin, ``{Conformal Anomaly in Weyl Theory and Anomaly Free Superconformal Theories},'' \href{http://dx.doi.org/10.1016/0370-2693(84)90668-3}{{\em Phys. Lett. B} {\bfseries 134} (1984) 187}.

\bibitem{Salvio:2017qkx}
A.~Salvio and A.~Strumia, ``{Agravity up to infinite energy},'' \href{http://dx.doi.org/10.1140/epjc/s10052-018-5588-4}{{\em Eur. Phys. J. C} {\bfseries 78} no.~2, (2018) 124}, \href{http://arxiv.org/abs/1705.03896}{{\ttfamily arXiv:1705.03896 [hep-th]}}.

\bibitem{Percacci:2023rbo}
R.~Percacci, ``{Gravity as a Quantum Field Theory},'' \href{http://dx.doi.org/10.3390/sym15020449}{{\em Symmetry} {\bfseries 15} no.~2, (2023) 449}.

\bibitem{Melichev:2023lwj}
O.~Melichev and R.~Percacci, ``{On the renormalization of Poincar\'e gauge theories},'' \href{http://dx.doi.org/10.1007/JHEP03(2024)133}{{\em JHEP} {\bfseries 03} (2024) 133}, \href{http://arxiv.org/abs/2307.02336}{{\ttfamily arXiv:2307.02336 [hep-th]}}.

\bibitem{Herrero-Valea:2023zex}
M.~Herrero-Valea, ``{The status of Ho\v{r}ava gravity},'' \href{http://dx.doi.org/10.1140/epjp/s13360-023-04593-y}{{\em Eur. Phys. J. Plus} {\bfseries 138} no.~11, (2023) 968}, \href{http://arxiv.org/abs/2307.13039}{{\ttfamily arXiv:2307.13039 [gr-qc]}}.

\bibitem{Anselmi:2018cau}
A.~Damiano, ``The correspondence principle in quantum field theory and quantum gravity,'' September, 2018.
\newblock \url{https://philsci-archive.pitt.edu/15287/}.

\bibitem{Asorey:1996hz}
M.~Asorey, J.~L. Lopez, and I.~L. Shapiro, ``{Some remarks on high derivative quantum gravity},'' \href{http://dx.doi.org/10.1142/S0217751X97002991}{{\em Int. J. Mod. Phys.} {\bfseries A12} (1997) 5711--5734},
\href{http://arxiv.org/abs/hep-th/9610006}{{\ttfamily arXiv:hep-th/9610006 [hep-th]}}.

\bibitem{Modesto:2015ozb}
L.~Modesto and I.~L. Shapiro, ``{Superrenormalizable quantum gravity with complex ghosts},'' \href{http://dx.doi.org/10.1016/j.physletb.2016.02.021}{{\em Phys. Lett.} {\bfseries B755} (2016) 279--284},
\href{http://arxiv.org/abs/1512.07600}{{\ttfamily arXiv:1512.07600 [hep-th]}}.

\bibitem{Modesto:2011kw}
L.~Modesto, ``{Super-renormalizable Quantum Gravity},'' \href{http://dx.doi.org/10.1103/PhysRevD.86.044005}{{\em Phys. Rev.} {\bfseries D86} (2012) 044005},
\href{http://arxiv.org/abs/1107.2403}{{\ttfamily arXiv:1107.2403 [hep-th]}}.

\bibitem{Koshelev:2017ebj}
A.~S. Koshelev, K.~Sravan~Kumar, L.~Modesto, and L.~Rachwał, ``{Finite quantum gravity in dS and AdS spacetimes},'' \href{http://dx.doi.org/10.1103/PhysRevD.98.046007}{{\em Phys. Rev.} {\bfseries D98} no.~4, (2018) 046007},
\href{http://arxiv.org/abs/1710.07759}{{\ttfamily arXiv:1710.07759 [hep-th]}}.

\bibitem{BasiBeneito:2022wux}
A.~Bas~i Beneito, G.~Calcagni, and L.~Rachwa\l{}, {\em {Classical and Quantum Nonlocal Gravity}}.
\newblock 2024.
\newblock \href{http://arxiv.org/abs/2211.05606}{{\ttfamily arXiv:2211.05606 [hep-th]}}.

\bibitem{Buoninfante:2021xxl}
L.~Buoninfante, ``{Maximal acceleration, reciprocity \& nonlocality},'' \href{http://dx.doi.org/10.1142/S0218271821420128}{{\em Int. J. Mod. Phys. D} {\bfseries 30} no.~14, (2021) 2142012}, \href{http://arxiv.org/abs/2105.08167}{{\ttfamily arXiv:2105.08167 [hep-th]}}.

\bibitem{Buoninfante:2022ykf}
L.~Buoninfante and K.~S. Kumar, ``{Quantum gravity, higher derivatives and nonlocality},'' \href{http://dx.doi.org/10.1393/ncc/i2022-22025-8}{{\em Nuovo Cim. C} {\bfseries 45} no.~2, (2022) 25}.

\bibitem{Buoninfante:2024yth}
L.~Buoninfante {\em et~al.}, ``{Visions in Quantum Gravity},'' \href{http://arxiv.org/abs/2412.08696}{{\ttfamily arXiv:2412.08696 [hep-th]}}.

\bibitem{Woodard:2015zca}
R.~P. Woodard, ``{Ostrogradsky's theorem on Hamiltonian instability},'' \href{http://dx.doi.org/10.4249/scholarpedia.32243}{{\em Scholarpedia} {\bfseries 10} no.~8, (2015) 32243},
\href{http://arxiv.org/abs/1506.02210}{{\ttfamily arXiv:1506.02210 [hep-th]}}.

\bibitem{Held:2023aap}
A.~Held and H.~Lim, ``{Nonlinear evolution of quadratic gravity in 3+1 dimensions},'' \href{http://dx.doi.org/10.1103/PhysRevD.108.104025}{{\em Phys. Rev. D} {\bfseries 108} no.~10, (2023) 104025}, \href{http://arxiv.org/abs/2306.04725}{{\ttfamily arXiv:2306.04725 [gr-qc]}}.

\bibitem{Held:2025ckb}
A.~Held and H.~Lim, ``{Black-hole binaries and waveforms in Quadratic Gravity},'' \href{http://arxiv.org/abs/2503.13428}{{\ttfamily arXiv:2503.13428 [gr-qc]}}.

\bibitem{Deffayet:2021nnt}
C.~Deffayet, S.~Mukohyama, and A.~Vikman, ``{Ghosts without Runaway Instabilities},'' \href{http://dx.doi.org/10.1103/PhysRevLett.128.041301}{{\em Phys. Rev. Lett.} {\bfseries 128} no.~4, (2022) 041301}, \href{http://arxiv.org/abs/2108.06294}{{\ttfamily arXiv:2108.06294 [gr-qc]}}.

\bibitem{Deffayet:2023wdg}
C.~Deffayet, A.~Held, S.~Mukohyama, and A.~Vikman, ``{Global and local stability for ghosts coupled to positive energy degrees of freedom},'' \href{http://dx.doi.org/10.1088/1475-7516/2023/11/031}{{\em JCAP} {\bfseries 11} (2023) 031}, \href{http://arxiv.org/abs/2305.09631}{{\ttfamily arXiv:2305.09631 [gr-qc]}}.

\bibitem{ErrastiDiez:2024hfq}
V.~Errasti~D\'\i{}ez, J.~Gaset~Rif\`a, and G.~Staudt, ``{Foundations of ghost stability},'' \href{http://arxiv.org/abs/2408.16832}{{\ttfamily arXiv:2408.16832 [hep-th]}}.

\bibitem{Gross:2020tph}
C.~Gross, A.~Strumia, D.~Teresi, and M.~Zirilli, ``{Is negative kinetic energy metastable?},'' \href{http://dx.doi.org/10.1103/PhysRevD.103.115025}{{\em Phys. Rev. D} {\bfseries 103} no.~11, (2021) 115025}, \href{http://arxiv.org/abs/2007.05541}{{\ttfamily arXiv:2007.05541 [hep-th]}}.

\bibitem{Donoghue:2021eto}
J.~F. Donoghue and G.~Menezes, ``{Ostrogradsky instability can be overcome by quantum physics},'' \href{http://dx.doi.org/10.1103/PhysRevD.104.045010}{{\em Phys. Rev. D} {\bfseries 104} no.~4, (2021) 045010}, \href{http://arxiv.org/abs/2105.00898}{{\ttfamily arXiv:2105.00898 [hep-th]}}.

\bibitem{Woodard:2023tgb}
R.~P. Woodard, ``{Don\textquoteright{}t throw the baby out with the bath water},'' \href{http://dx.doi.org/10.1140/epjp/s13360-023-04709-4}{{\em Eur. Phys. J. Plus} {\bfseries 138} no.~11, (2023) 1067}, \href{http://arxiv.org/abs/2306.09596}{{\ttfamily arXiv:2306.09596 [gr-qc]}}.

\bibitem{Kubo:2023lpz}
J.~Kubo and T.~Kugo, ``{Unitarity violation in field theories of Lee\textendash{}Wick\textquoteright{}s complex ghost},'' \href{http://dx.doi.org/10.1093/ptep/ptad143}{{\em PTEP} {\bfseries 2023} no.~12, (2023) 123B02}, \href{http://arxiv.org/abs/2308.09006}{{\ttfamily arXiv:2308.09006 [hep-th]}}.

\bibitem{Salvio:2015gsi}
A.~Salvio and A.~Strumia, ``{Quantum mechanics of 4-derivative theories},'' \href{http://dx.doi.org/10.1140/epjc/s10052-016-4079-8}{{\em Eur. Phys. J. C} {\bfseries 76} no.~4, (2016) 227}, \href{http://arxiv.org/abs/1512.01237}{{\ttfamily arXiv:1512.01237 [hep-th]}}.

\bibitem{Strumia:2017dvt}
A.~Strumia, ``{Interpretation of quantum mechanics with indefinite norm},'' \href{http://dx.doi.org/10.3390/physics1010003}{{\em MDPI Physics} {\bfseries 1} no.~1, (2019) 17--32}, \href{http://arxiv.org/abs/1709.04925}{{\ttfamily arXiv:1709.04925 [quant-ph]}}.

\bibitem{Kuntz:2024rzu}
J.~Kuntz, ``{Unitarity through PT symmetry in Quantum Quadratic Gravity},'' \href{http://arxiv.org/abs/2410.08278}{{\ttfamily arXiv:2410.08278 [hep-th]}}.

\bibitem{Holdom:2024onr}
B.~Holdom, ``{Making sense of ghosts},'' \href{http://dx.doi.org/10.1016/j.nuclphysb.2024.116696}{{\em Nucl. Phys. B} {\bfseries 1008} (2024) 116696}, \href{http://arxiv.org/abs/2408.04089}{{\ttfamily arXiv:2408.04089 [hep-th]}}.

\bibitem{Feynman1987-FEYNP}
R.~P. Feynman, ``Negative probability,'' in {\em Quantum Implications: Essays in Honour of David Bohm}, B.~J. Hiley and D.~Peat, eds., pp.~235--248.
\newblock Methuen, 1987.

\bibitem{Coleman:1969xz}
S.~Coleman, ``{Acausality},'' in {\em {7th International School of Subnuclear Physics (Ettore Majorana): Subnuclear Phenomena}}.
\newblock 1969.

\bibitem{Kubo:2024ysu}
J.~Kubo and T.~Kugo, ``{Anti-Instability of Complex Ghost},'' \href{http://dx.doi.org/10.1093/ptep/ptae053}{{\em PTEP} {\bfseries 2024} no.~5, (2024) 053B01}, \href{http://arxiv.org/abs/2402.15956}{{\ttfamily arXiv:2402.15956 [hep-th]}}.

\bibitem{Buoninfante:2025klm}
L.~Buoninfante, ``{Remarks on ghost resonances},'' \href{http://dx.doi.org/10.1007/JHEP02(2025)186}{{\em JHEP} {\bfseries 02} (2025) 186}, \href{http://arxiv.org/abs/2501.04097}{{\ttfamily arXiv:2501.04097 [hep-th]}}.

\bibitem{Anselmi:2023wjx}
D.~Anselmi, ``{Propagators and widths of physical and purely virtual particles in a finite interval of time},'' \href{http://dx.doi.org/10.1007/JHEP07(2023)099}{{\em JHEP} {\bfseries 07} (2023) 099}, \href{http://arxiv.org/abs/2304.07643}{{\ttfamily arXiv:2304.07643 [hep-ph]}}.

\bibitem{Grinstein:2008bg}
B.~Grinstein, D.~O'Connell, and M.~B. Wise, ``{Causality as an emergent macroscopic phenomenon: The Lee-Wick O(N) model},'' \href{http://dx.doi.org/10.1103/PhysRevD.79.105019}{{\em Phys. Rev. D} {\bfseries 79} (2009) 105019}, \href{http://arxiv.org/abs/0805.2156}{{\ttfamily arXiv:0805.2156 [hep-th]}}.

\bibitem{Anselmi:2021hab}
D.~Anselmi, ``{Diagrammar of physical and fake particles and spectral optical theorem},'' \href{http://dx.doi.org/10.1007/JHEP11(2021)030}{{\em JHEP} {\bfseries 11} (2021) 030}, \href{http://arxiv.org/abs/2109.06889}{{\ttfamily arXiv:2109.06889 [hep-th]}}.

\bibitem{Rickles:2020zwh}
D.~Rickles, {\em {Covered with Deep Mist}}.
\newblock Oxford University Press, 3, 2020.

\bibitem{Basile:2025zjc}
I.~Basile, B.~Knorr, A.~Platania, and M.~Schiffer, ``{Asymptotic safety, quantum gravity, and the swampland: a conceptual assessment},'' \href{http://arxiv.org/abs/2502.12290}{{\ttfamily arXiv:2502.12290 [hep-th]}}.

\bibitem{Loll:2022ibq}
R.~Loll, G.~Fabiano, D.~Frattulillo, and F.~Wagner, ``{Quantum Gravity in 30 Questions},'' \href{http://dx.doi.org/10.22323/1.406.0316}{{\em PoS} {\bfseries CORFU2021} (2022) 316}, \href{http://arxiv.org/abs/2206.06762}{{\ttfamily arXiv:2206.06762 [hep-th]}}.

\bibitem{Planck:2018jri}
{\bfseries Planck} Collaboration, Y.~Akrami {\em et~al.}, ``{Planck 2018 results. X. Constraints on inflation},'' \href{http://dx.doi.org/10.1051/0004-6361/201833887}{{\em Astron. Astrophys.} {\bfseries 641} (2020) A10}, \href{http://arxiv.org/abs/1807.06211}{{\ttfamily arXiv:1807.06211 [astro-ph.CO]}}.

\bibitem{Planck:2018vyg}
{\bfseries Planck} Collaboration, N.~Aghanim {\em et~al.}, ``{Planck 2018 results. VI. Cosmological parameters},'' \href{http://dx.doi.org/10.1051/0004-6361/201833910}{{\em Astron. Astrophys.} {\bfseries 641} (2020) A6}, \href{http://arxiv.org/abs/1807.06209}{{\ttfamily arXiv:1807.06209 [astro-ph.CO]}}. [Erratum: Astron.Astrophys. 652, C4 (2021)].

\bibitem{Martin:2013tda}
J.~Martin, C.~Ringeval, and V.~Vennin, ``{Encyclopædia Inflationaris},'' \href{http://dx.doi.org/10.1016/j.dark.2014.01.003}{{\em Phys. Dark Univ.} {\bfseries 5-6} (2014) 75--235},
\href{http://arxiv.org/abs/1303.3787}{{\ttfamily arXiv:1303.3787 [astro-ph.CO]}}.

\bibitem{Starobinsky:1980te}
A.~A. Starobinsky, ``{A New Type of Isotropic Cosmological Models Without Singularity},'' \href{http://dx.doi.org/10.1016/0370-2693(80)90670-X}{{\em Phys. Lett. B} {\bfseries 91} (1980) 99--102}.

\bibitem{Mukhanov:1981xt}
V.~F. Mukhanov and G.~V. Chibisov, ``{Quantum Fluctuations and a Nonsingular Universe},'' {\em JETP Lett.} {\bfseries 33} (1981) 532--535.
[Pisma Zh. Eksp. Teor. Fiz.33,549(1981)].

\bibitem{Starobinsky:1981vz}
A.~A. Starobinsky, ``{NONSINGULAR MODEL OF THE UNIVERSE WITH THE QUANTUM GRAVITATIONAL DE SITTER STAGE AND ITS OBSERVATIONAL CONSEQUENCES},'' in {\em {*Moscow 1981, Proceedings of the second seminar ``Quantum Theory of Gravity''*, 58-72}}.
\newblock
1981.
\newblock

\bibitem{Starobinsky:1983zz}
A.~A. Starobinsky, ``{The Perturbation Spectrum Evolving from a Nonsingular Initially De-Sitter Cosmology and the Microwave Background Anisotropy},''
{\em Sov. Astron. Lett.} {\bfseries 9} (1983) 302.

\bibitem{Kofman:1985aw}
L.~A. Kofman, A.~D. Linde, and A.~A. Starobinsky, ``{Inflationary Universe Generated by the Combined Action of a Scalar Field and Gravitational Vacuum Polarization},'' \href{http://dx.doi.org/10.1016/0370-2693(85)90381-8}{{\em Phys. Lett. B} {\bfseries 157} (1985) 361--367}.

\bibitem{Vilenkin:1985md}
A.~Vilenkin, ``{Classical and Quantum Cosmology of the Starobinsky Inflationary Model},'' \href{http://dx.doi.org/10.1103/PhysRevD.32.2511}{{\em Phys. Rev. D} {\bfseries 32} (1985) 2511}.

\bibitem{Ellis:2023wic}
J.~Ellis and D.~Wands, ``{Inflation (2023)},'' \href{http://arxiv.org/abs/2312.13238}{{\ttfamily arXiv:2312.13238 [astro-ph.CO]}}.

\bibitem{Ketov:2025nkr}
S.~V. Ketov, ``{On Legacy of Starobinsky Inflation},''
\newblock 1, 2025.
\newblock \href{http://arxiv.org/abs/2501.06451}{{\ttfamily arXiv:2501.06451 [gr-qc]}}.

\bibitem{ACT:2025fju}
{\bfseries ACT} Collaboration, T.~Louis {\em et~al.}, ``{The Atacama Cosmology Telescope: DR6 Power Spectra, Likelihoods and $\Lambda$CDM Parameters},'' \href{http://arxiv.org/abs/2503.14452}{{\ttfamily arXiv:2503.14452 [astro-ph.CO]}}.

\bibitem{ACT:2025tim}
{\bfseries ACT} Collaboration, E.~Calabrese {\em et~al.}, ``{The Atacama Cosmology Telescope: DR6 Constraints on Extended Cosmological Models},'' \href{http://arxiv.org/abs/2503.14454}{{\ttfamily arXiv:2503.14454 [astro-ph.CO]}}.

\bibitem{DESI:2024uvr}
{\bfseries DESI} Collaboration, A.~G. Adame {\em et~al.}, ``{DESI 2024 III: Baryon Acoustic Oscillations from Galaxies and Quasars},'' \href{http://arxiv.org/abs/2404.03000}{{\ttfamily arXiv:2404.03000 [astro-ph.CO]}}.

\bibitem{DESI:2024mwx}
{\bfseries DESI} Collaboration, A.~G. Adame {\em et~al.}, ``{DESI 2024 VI: cosmological constraints from the measurements of baryon acoustic oscillations},'' \href{http://dx.doi.org/10.1088/1475-7516/2025/02/021}{{\em JCAP} {\bfseries 02} (2025) 021}, \href{http://arxiv.org/abs/2404.03002}{{\ttfamily arXiv:2404.03002 [astro-ph.CO]}}.

\bibitem{SPT-3G:2024atg}
{\bfseries SPT-3G} Collaboration, F.~Ge {\em et~al.}, ``{Cosmology from CMB lensing and delensed EE power spectra using 2019\textendash{}2020 SPT-3G polarization data},'' \href{http://dx.doi.org/10.1103/PhysRevD.111.083534}{{\em Phys. Rev. D} {\bfseries 111} no.~8, (2025) 083534}, \href{http://arxiv.org/abs/2411.06000}{{\ttfamily arXiv:2411.06000 [astro-ph.CO]}}.

\bibitem{Gorbunov:2010bn}
D.~S. Gorbunov and A.~G. Panin, ``{Scalaron the mighty: producing dark matter and baryon asymmetry at reheating},'' \href{http://dx.doi.org/10.1016/j.physletb.2011.04.067}{{\em Phys. Lett. B} {\bfseries 700} (2011) 157--162}, \href{http://arxiv.org/abs/1009.2448}{{\ttfamily arXiv:1009.2448 [hep-ph]}}.

\bibitem{Jeong:2023zrv}
H.~Jeong, K.~Kamada, A.~A. Starobinsky, and J.~Yokoyama, ``{Reheating process in the R $^{2}$ inflationary model with the baryogenesis scenario},'' \href{http://dx.doi.org/10.1088/1475-7516/2023/11/023}{{\em JCAP} {\bfseries 11} (2023) 023}, \href{http://arxiv.org/abs/2305.14273}{{\ttfamily arXiv:2305.14273 [hep-ph]}}.

\bibitem{del-Corral:2023apl}
D.~del Corral, P.~Gondolo, K.~S. Kumar, and J.~a. Marto, ``{Revisiting primordial black holes formation from preheating instabilities: the case of Starobinsky inflation},'' \href{http://dx.doi.org/10.1088/1475-7516/2025/02/009}{{\em JCAP} {\bfseries 02} (2025) 009}, \href{http://arxiv.org/abs/2311.02754}{{\ttfamily arXiv:2311.02754 [astro-ph.CO]}}.

\bibitem{Garcia:2023tkk}
M.~A.~G. Garcia, G.~Germ\'an, R.~Gonzalez~Quaglia, and A.~M.~M. Colorado, ``{Reheating constraints and consistency relations of the Starobinsky model and some of its generalizations},'' \href{http://dx.doi.org/10.1088/1475-7516/2023/12/015}{{\em JCAP} {\bfseries 12} (2023) 015}, \href{http://arxiv.org/abs/2306.15831}{{\ttfamily arXiv:2306.15831 [astro-ph.CO]}}.

\bibitem{Dorsch:2024nan}
G.~C. Dorsch, L.~Miranda, and N.~Yokomizo, ``{Gravitational reheating in Starobinsky inflation},'' \href{http://dx.doi.org/10.1088/1475-7516/2024/11/050}{{\em JCAP} {\bfseries 11} (2024) 050}, \href{http://arxiv.org/abs/2406.04161}{{\ttfamily arXiv:2406.04161 [gr-qc]}}.

\bibitem{Deruelle:2012xv}
N.~Deruelle, M.~Sasaki, Y.~Sendouda, and A.~Youssef, ``{Lorentz-violating vs ghost gravitons: the example of Weyl gravity},'' \href{http://dx.doi.org/10.1007/JHEP09(2012)009}{{\em JHEP} {\bfseries 09} (2012) 009}, \href{http://arxiv.org/abs/1202.3131}{{\ttfamily arXiv:1202.3131 [gr-qc]}}.

\bibitem{Ivanov:2016hcm}
M.~M. Ivanov and A.~A. Tokareva, ``{Cosmology with a light ghost},'' \href{http://dx.doi.org/10.1088/1475-7516/2016/12/018}{{\em JCAP} {\bfseries 12} (2016) 018}, \href{http://arxiv.org/abs/1610.05330}{{\ttfamily arXiv:1610.05330 [hep-th]}}.

\bibitem{Salvio:2017xul}
A.~Salvio, ``{Inflationary Perturbations in No-Scale Theories},'' \href{http://dx.doi.org/10.1140/epjc/s10052-017-4825-6}{{\em Eur. Phys. J. C} {\bfseries 77} no.~4, (2017) 267}, \href{http://arxiv.org/abs/1703.08012}{{\ttfamily arXiv:1703.08012 [astro-ph.CO]}}.

\bibitem{Anselmi:2020lpp}
D.~Anselmi, E.~Bianchi, and M.~Piva, ``{Predictions of quantum gravity in inflationary cosmology: effects of the Weyl-squared term},'' \href{http://dx.doi.org/10.1007/JHEP07(2020)211}{{\em JHEP} {\bfseries 07} (2020) 211}, \href{http://arxiv.org/abs/2005.10293}{{\ttfamily arXiv:2005.10293 [hep-th]}}.

\bibitem{Salvio:2022mld}
A.~Salvio, ``{BICEP/Keck data and quadratic gravity},'' \href{http://dx.doi.org/10.1088/1475-7516/2022/09/027}{{\em JCAP} {\bfseries 09} (2022) 027}, \href{http://arxiv.org/abs/2202.00684}{{\ttfamily arXiv:2202.00684 [astro-ph.CO]}}.

\bibitem{Kubo:2025jla}
J.~Kubo and J.~Kuntz, ``{Primordial Gravitational Waves in Quadratic Gravity},'' \href{http://arxiv.org/abs/2502.03543}{{\ttfamily arXiv:2502.03543 [gr-qc]}}.

\bibitem{BICEP:2021xfz}
{\bfseries BICEP, Keck} Collaboration, P.~A.~R. Ade {\em et~al.}, ``{Improved Constraints on Primordial Gravitational Waves using Planck, WMAP, and BICEP/Keck Observations through the 2018 Observing Season},'' \href{http://dx.doi.org/10.1103/PhysRevLett.127.151301}{{\em Phys. Rev. Lett.} {\bfseries 127} no.~15, (2021) 151301}, \href{http://arxiv.org/abs/2110.00483}{{\ttfamily arXiv:2110.00483 [astro-ph.CO]}}.

\bibitem{Tristram:2021tvh}
M.~Tristram {\em et~al.}, ``{Improved limits on the tensor-to-scalar ratio using BICEP and Planck data},'' \href{http://dx.doi.org/10.1103/PhysRevD.105.083524}{{\em Phys. Rev. D} {\bfseries 105} no.~8, (2022) 083524}, \href{http://arxiv.org/abs/2112.07961}{{\ttfamily arXiv:2112.07961 [astro-ph.CO]}}.

\bibitem{CORE:2016ymi}
{\bfseries CORE} Collaboration, F.~Finelli {\em et~al.}, ``{Exploring cosmic origins with CORE: Inflation},'' \href{http://dx.doi.org/10.1088/1475-7516/2018/04/016}{{\em JCAP} {\bfseries 04} (2018) 016}, \href{http://arxiv.org/abs/1612.08270}{{\ttfamily arXiv:1612.08270 [astro-ph.CO]}}.

\bibitem{CMB-S4:2020lpa}
{\bfseries CMB-S4} Collaboration, K.~Abazajian {\em et~al.}, ``{CMB-S4: Forecasting Constraints on Primordial Gravitational Waves},'' \href{http://dx.doi.org/10.3847/1538-4357/ac1596}{{\em Astrophys. J.} {\bfseries 926} no.~1, (2022) 54}, \href{http://arxiv.org/abs/2008.12619}{{\ttfamily arXiv:2008.12619 [astro-ph.CO]}}.

\bibitem{Paoletti:2022kij}
{\bfseries LiteBIRD} Collaboration, D.~Paoletti, ``{The $LiteBIRD$ mission},'' \href{http://dx.doi.org/10.22323/1.414.0085}{{\em PoS} {\bfseries ICHEP2022} (11, 2022) 085}.

\bibitem{NASAPICO:2019thw}
{\bfseries NASA PICO} Collaboration, S.~Hanany {\em et~al.}, ``{PICO: Probe of Inflation and Cosmic Origins},'' \href{http://arxiv.org/abs/1902.10541}{{\ttfamily arXiv:1902.10541 [astro-ph.IM]}}.

\bibitem{SimonsObservatory:2018koc}
{\bfseries Simons Observatory} Collaboration, P.~Ade {\em et~al.}, ``{The Simons Observatory: Science goals and forecasts},'' \href{http://dx.doi.org/10.1088/1475-7516/2019/02/056}{{\em JCAP} {\bfseries 02} (2019) 056}, \href{http://arxiv.org/abs/1808.07445}{{\ttfamily arXiv:1808.07445 [astro-ph.CO]}}.

\bibitem{Euclid:2021qvm}
{\bfseries Euclid} Collaboration, S.~Ili\'c {\em et~al.}, ``{Euclid preparation. XV. Forecasting cosmological constraints for the Euclid and CMB joint analysis},'' \href{http://dx.doi.org/10.1051/0004-6361/202141556}{{\em Astron. Astrophys.} {\bfseries 657} (2022) A91}, \href{http://arxiv.org/abs/2106.08346}{{\ttfamily arXiv:2106.08346 [astro-ph.CO]}}.

\bibitem{Dona:2015tra}
P.~Donà, S.~Giaccari, L.~Modesto, L.~Rachwal, and Y.~Zhu, ``{Scattering amplitudes in super-renormalizable gravity},'' \href{http://dx.doi.org/10.1007/JHEP08(2015)038}{{\em JHEP} {\bfseries 08} (2015) 038},
\href{http://arxiv.org/abs/1506.04589}{{\ttfamily arXiv:1506.04589 [hep-th]}}.

\bibitem{Buccio:2023lzo}
D.~Buccio, J.~F. Donoghue, and R.~Percacci, ``{Amplitudes and Renormalization Group Techniques: A Case Study},'' \href{http://arxiv.org/abs/2307.00055}{{\ttfamily arXiv:2307.00055 [hep-th]}}.

\bibitem{Aydemir:2012nz}
U.~Aydemir, M.~M. Anber, and J.~F. Donoghue, ``{Self-healing of unitarity in effective field theories and the onset of new physics},'' \href{http://dx.doi.org/10.1103/PhysRevD.86.014025}{{\em Phys. Rev. D} {\bfseries 86} (2012) 014025}, \href{http://arxiv.org/abs/1203.5153}{{\ttfamily arXiv:1203.5153 [hep-ph]}}.

\bibitem{Piva:2023eaj}
M.~Piva, ``{High-Energy Behavior of Scattering Amplitudes in Theories with Purely Virtual Particles},'' \href{http://dx.doi.org/10.1007/JHEP05(2024)231}{{\em JHEP} {\bfseries 05} (2024) 231}, \href{http://arxiv.org/abs/2312.09045}{{\ttfamily arXiv:2312.09045 [hep-th]}}.

\bibitem{tHooft:1987vrq}
G.~'t~Hooft, ``{Graviton Dominance in Ultrahigh-Energy Scattering},'' \href{http://dx.doi.org/10.1016/0370-2693(87)90159-6}{{\em Phys. Lett. B} {\bfseries 198} (1987) 61--63}.

\bibitem{Giddings:2007qq}
S.~B. Giddings and M.~Srednicki, ``{High-energy gravitational scattering and black hole resonances},'' \href{http://dx.doi.org/10.1103/PhysRevD.77.085025}{{\em Phys. Rev. D} {\bfseries 77} (2008) 085025}, \href{http://arxiv.org/abs/0711.5012}{{\ttfamily arXiv:0711.5012 [hep-th]}}.

\bibitem{Dvali:2014ila}
G.~Dvali, C.~Gomez, R.~S. Isermann, D.~L\"ust, and S.~Stieberger, ``{Black hole formation and classicalization in ultra-Planckian 2\textrightarrow{}N scattering},'' \href{http://dx.doi.org/10.1016/j.nuclphysb.2015.02.004}{{\em Nucl. Phys. B} {\bfseries 893} (2015) 187--235}, \href{http://arxiv.org/abs/1409.7405}{{\ttfamily arXiv:1409.7405 [hep-th]}}.

\bibitem{Addazi:2016ksu}
A.~Addazi, M.~Bianchi, and G.~Veneziano, ``{Glimpses of black hole formation/evaporation in highly inelastic, ultra-planckian string collisions},'' \href{http://dx.doi.org/10.1007/JHEP02(2017)111}{{\em JHEP} {\bfseries 02} (2017) 111}, \href{http://arxiv.org/abs/1611.03643}{{\ttfamily arXiv:1611.03643 [hep-th]}}.

\bibitem{Lu:2015cqa}
H.~Lu, A.~Perkins, C.~N. Pope, and K.~S. Stelle, ``{Black Holes in Higher-Derivative Gravity},'' \href{http://dx.doi.org/10.1103/PhysRevLett.114.171601}{{\em Phys. Rev. Lett.} {\bfseries 114} no.~17, (2015) 171601},
\href{http://arxiv.org/abs/1502.01028}{{\ttfamily arXiv:1502.01028 [hep-th]}}.

\bibitem{Lu:2015psa}
H.~Lü, A.~Perkins, C.~N. Pope, and K.~S. Stelle, ``{Spherically Symmetric Solutions in Higher-Derivative Gravity},'' \href{http://dx.doi.org/10.1103/PhysRevD.92.124019}{{\em Phys. Rev.} {\bfseries D92} no.~12, (2015) 124019},
\href{http://arxiv.org/abs/1508.00010}{{\ttfamily arXiv:1508.00010 [hep-th]}}.

\bibitem{Buoninfante:2024oyi}
L.~Buoninfante, F.~Di~Filippo, I.~Kol\'a\v{r}, and F.~Saueressig, ``{Dust collapse and horizon formation in quadratic gravity},'' \href{http://dx.doi.org/10.1088/1475-7516/2025/01/114}{{\em JCAP} {\bfseries 01} (2025) 114}, \href{http://arxiv.org/abs/2410.05941}{{\ttfamily arXiv:2410.05941 [gr-qc]}}.

\bibitem{Holdom:2016nek}
B.~Holdom and J.~Ren, ``{Not quite a black hole},'' \href{http://dx.doi.org/10.1103/PhysRevD.95.084034}{{\em Phys. Rev.} {\bfseries D95} no.~8, (2017) 084034},
\href{http://arxiv.org/abs/1612.04889}{{\ttfamily arXiv:1612.04889 [gr-qc]}}.

\bibitem{Jaffe:1966an}
A.~M. Jaffe, ``{HIGH-ENERGY BEHAVIOR OF LOCAL QUANTUM FIELDS},'' {\em SLAC PUB 0250, Stanford Linear Accelerator Center} (12, 1966) .

\bibitem{Jaffe:1967nb}
A.~M. Jaffe, ``{HIGH-ENERGY BEHAVIOR IN QUANTUM FIELD THEORY. I. STRICTLY LOCALIZABLE FIELDS},'' \href{http://dx.doi.org/10.1103/PhysRev.158.1454}{{\em Phys. Rev.} {\bfseries 158} (1967) 1454--1461}.

\bibitem{Keltner:2015xda}
L.~Keltner and A.~J. Tolley, ``{UV properties of Galileons: Spectral Densities},'' \href{http://arxiv.org/abs/1502.05706}{{\ttfamily arXiv:1502.05706 [hep-th]}}.

\bibitem{Buoninfante:2023dyd}
L.~Buoninfante, J.~Tokuda, and M.~Yamaguchi, ``{New lower bounds on scattering amplitudes: non-locality constraints},'' \href{http://dx.doi.org/10.1007/JHEP01(2024)082}{{\em JHEP} {\bfseries 01} (2024) 082}, \href{http://arxiv.org/abs/2305.16422}{{\ttfamily arXiv:2305.16422 [hep-th]}}.

\bibitem{Buoninfante:2024ibt}
L.~Buoninfante, L.-Q. Shao, and A.~Tokareva, ``{Non-local positivity bounds: islands in Terra Incognita},'' \href{http://arxiv.org/abs/2412.08634}{{\ttfamily arXiv:2412.08634 [hep-th]}}.

\bibitem{Vafa:2005ui}
C.~Vafa, ``{The String landscape and the swampland},'' \href{http://arxiv.org/abs/hep-th/0509212}{{\ttfamily arXiv:hep-th/0509212}}.

\bibitem{Basile:2021krr}
I.~Basile and A.~Platania, ``{Asymptotic Safety: Swampland or Wonderland?},'' \href{http://dx.doi.org/10.3390/universe7100389}{{\em Universe} {\bfseries 7} no.~10, (2021) 389}, \href{http://arxiv.org/abs/2107.06897}{{\ttfamily arXiv:2107.06897 [hep-th]}}.

\bibitem{Eichhorn:2024rkc}
A.~Eichhorn, A.~Hebecker, J.~M. Pawlowski, and J.~Walcher, ``{The absolute swampland},'' \href{http://dx.doi.org/10.1209/0295-5075/ada1f3}{{\em EPL} {\bfseries 149} no.~3, (2025) 39001}, \href{http://arxiv.org/abs/2405.20386}{{\ttfamily arXiv:2405.20386 [hep-th]}}.

\bibitem{Lust:2023zql}
D.~L\"ust, J.~Masias, B.~Muntz, and M.~Scalisi, ``{Starobinsky inflation in the swampland},'' \href{http://dx.doi.org/10.1007/JHEP07(2024)186}{{\em JHEP} {\bfseries 07} (2024) 186}, \href{http://arxiv.org/abs/2312.13210}{{\ttfamily arXiv:2312.13210 [hep-th]}}.

\bibitem{Brinkmann:2023eph}
M.~Brinkmann, M.~Cicoli, and P.~Zito, ``{Starobinsky inflation from string theory?},'' \href{http://dx.doi.org/10.1007/JHEP09(2023)038}{{\em JHEP} {\bfseries 09} (2023) 038}, \href{http://arxiv.org/abs/2305.05703}{{\ttfamily arXiv:2305.05703 [hep-th]}}.

\bibitem{Weinberg-history-qft}
S.~Weinberg, ``The search for unity: Notes for a history of quantum field theory,'' {\em Daedalus} {\bfseries 106} no.~4, (1977) 17--35. \url{http://www.jstor.org/stable/20024506}.

\bibitem{umezawa-qft}
H.~Umezawa, ``Development in concepts in quantum field theory in half century,'' {\em Math Japonica} {\bfseries 41} no.~1, (1995) 109--124. \url{https://drive.google.com/file/d/1iHCEYNFf8lPyWWb1Hbo-AsmIurpWBSPG/view?usp=sharing}.

\bibitem{tHooft:2015sdj}
G.~'t~Hooft, ``{The Evolution of Quantum Field Theory, From QED to Grand Unification},'' \href{http://dx.doi.org/10.1142/9789814733519_0001}{{\em Adv. Ser. Direct. High Energy Phys.} {\bfseries 26} (2016) 1--27}, \href{http://arxiv.org/abs/1503.05007}{{\ttfamily arXiv:1503.05007 [hep-th]}}.

\bibitem{Pais1986-PAIIBO}
A.~Pais, {\em Inward Bound: Of Matter and Forces in the Physical World}.
\newblock Oxford University Press, New York, 1986.

\bibitem{Brown:1993kb}
L.~M. Brown, ed., \href{http://dx.doi.org/10.1007/978-1-4612-2720-5}{{\em {Renormalization: From Lorentz to Landau (and beyond)}}}.
\newblock Springer New York, NY, 1993.

\bibitem{Hoddeson_Brown_Riordan_Dresden_1997}
L.~Brown, M.~Dresden, L.~Hoddeson, and M.~Riordan, \href{http://dx.doi.org/10.1017/CBO9780511471094.003}{{\em The Rise of the Standard Model: A History of Particle Physics from 1964 to 1979}}.
\newblock Cambridge University Press, 1997.

\bibitem{Cao:1997my}
T.~Y. Cao, \href{http://dx.doi.org/10.1017/9781108566926}{{\em {Conceptual developments of 20th century field theories}}}.
\newblock Cambridge University Press, 10, 2019.

\bibitem{galison:1987}
P.~Galison, {\em How Experiments End}.
\newblock University of Chicago Press, Chicago, 1987.

\bibitem{kaiser-drawing}
D.~Kaiser, {\em Drawing Theories Apart: The Dispersion of Feynman Diagrams in Postwar Physics}.
\newblock University of Chicago Press, 2005.

\bibitem{Oppenheimer:1930zz}
J.~R. Oppenheimer, ``{Note on the Theory of the Interaction of Field and Matter},'' \href{http://dx.doi.org/10.1103/PhysRev.35.461}{{\em Phys. Rev.} {\bfseries 35} (1930) 461--477}.

\bibitem{Dirac:1931kp}
P.~A.~M. Dirac, ``{Quantised singularities in the electromagnetic field,},'' \href{http://dx.doi.org/10.1098/rspa.1931.0130}{{\em Proc. Roy. Soc. Lond. A} {\bfseries 133} no.~821, (1931) 60--72}.

\bibitem{Heisenberg-minimal-length}
W.~Heisenberg, ``{\"Uber die in der Theorie der Elementarteilchen auftretende universelle L\"ange},'' \href{http://dx.doi.org/10.1002/andp.19384240105}{{\em Annalen der Physik} {\bfseries 32} (1938) 20}.

\bibitem{Heisenberg-S-matrix}
W.~Heisenberg, ``{Die beobachtbaren Größen in der Theorie der Elementarteilchen},'' \href{http://dx.doi.org/10.1007/BF01329800}{{\em Zeitschrift für Physik} {\bfseries 120} (1943) 513–538}.

\bibitem{Wheeler:1937zz}
J.~A. Wheeler, ``{On the Mathematical Description of Light Nuclei by the Method of Resonating Group Structure},'' \href{http://dx.doi.org/10.1103/PhysRev.52.1107}{{\em Phys. Rev.} {\bfseries 52} (1937) 1107--1122}.

\bibitem{dirac-neg-norm}
P.~A.~M. Dirac, ``The physical interpretation of quantum mechanics,'' \href{http://dx.doi.org/10.1098/rspa.1942.0023}{{\em Proc. Roy. Soc. Lond. A} {\bfseries 180} no.~980, (1942) 1--40}.

\bibitem{Born:1938zve}
M.~Born, ``{A suggestion for unifying quantum theory and relativity},'' \href{http://dx.doi.org/10.1098/rspa.1938.0060}{{\em Proc. Roy. Soc. Lond. A} {\bfseries 165} no.~921, (1938) 291--303}.

\bibitem{Born1}
M.~Born, ``{Reciprocity Theory of Elementary Particles},'' \href{http://dx.doi.org/10.1103/RevModPhys.21.463}{{\em Rev. Mod. Phys.} {\bfseries 21} (1949) 46}.

\bibitem{Born3}
M.~Born, ``{Reciprocity Theory of Electrodynamics},'' \href{http://dx.doi.org/10.1038/164281b0}{{\em Nature} {\bfseries 164} (1949) 281--282}.

\bibitem{Born:1950iuq}
M.~Born, ``{Non-Localizable Fields and Reciprocity},'' \href{http://dx.doi.org/10.1038/165269a0}{{\em Nature} {\bfseries 165} no.~4190, (1950) 269--270}.

\bibitem{Yukawa:1950eq}
H.~Yukawa, ``{Quantum Theory of Nonlocal Fields. 1. Free Fields},''
\href{http://dx.doi.org/10.1103/PhysRev.77.219}{{\em Phys. Rev.} {\bfseries 77} (1950) }.

\bibitem{Yukawa:1950er}
H.~Yukawa, ``{Quantum Theory of Nonlocal Fields. 2: Irreducible Fields and Their Interaction},''
\href{http://dx.doi.org/10.1103/PhysRev.80.1047}{{\em Phys. Rev.} {\bfseries 80} (1950) }.

\bibitem{Lamb:1947zz}
W.~E. Lamb and R.~C. Retherford, ``{Fine Structure of the Hydrogen Atom by a Microwave Method},'' \href{http://dx.doi.org/10.1103/PhysRev.72.241}{{\em Phys. Rev.} {\bfseries 72} (1947) 241--243}.

\bibitem{Bethe:1947id}
H.~A. Bethe, ``{The Electromagnetic shift of energy levels},'' \href{http://dx.doi.org/10.1103/PhysRev.72.339}{{\em Phys. Rev.} {\bfseries 72} (1947) 339--341}.

\bibitem{Heisenberg:1938xla}
W.~Heisenberg, ``{Die Grenzen der Anwendbarkeit der bisherigen Quantentheorie},'' \href{http://dx.doi.org/10.1007/BF01342872}{{\em Z. Phys.} {\bfseries 110} no.~3-4, (1938) 251--266}.

\bibitem{Bjorken:1968dy}
J.~D. Bjorken, ``{Asymptotic Sum Rules at Infinite Momentum},'' \href{http://dx.doi.org/10.1103/PhysRev.179.1547}{{\em Phys. Rev.} {\bfseries 179} (1969) 1547--1553}.

\bibitem{Bjorken:1969ja}
J.~D. Bjorken and E.~A. Paschos, ``{Inelastic Electron Proton and gamma Proton Scattering, and the Structure of the Nucleon},'' \href{http://dx.doi.org/10.1103/PhysRev.185.1975}{{\em Phys. Rev.} {\bfseries 185} (1969) 1975--1982}.

\bibitem{Feynman:1969wa}
R.~P. Feynman, ``{The behavior of hadron collisions at extreme energies},'' {\em Conf. Proc. C} {\bfseries 690905} (1969) 237--258.

\bibitem{Feynman:1969ej}
R.~P. Feynman, ``{Very high-energy collisions of hadrons},'' \href{http://dx.doi.org/10.1103/PhysRevLett.23.1415}{{\em Phys. Rev. Lett.} {\bfseries 23} (1969) }.

\bibitem{Sutherland:1967vf}
D.~G. Sutherland, ``{Current algebra and some nonstrong mesonic decays},'' \href{http://dx.doi.org/10.1016/0550-3213(67)90180-0}{{\em Nucl. Phys. B} {\bfseries 2} (1967) 433--440}.

\bibitem{veltman-paradox}
M.~J.~G. Veltman, ``{Theoretical aspects of high energy neutrino interactions},'' \href{http://dx.doi.org/10.1098/rspa.1967.0193}{{\em Proc. Roy. Soc. Lond. A} {\bfseries 301} (1967) 107}.

\bibitem{Landau:1958jby}
L.~D. Landau, A.~A. Abrikosov, A.~D. Galanin, L.~P. Gorkov, I.~Y. Pomeranchuk, and K.~A. Ter-Martirosyan, ``{Possibility of Formulation of a Theory of Strongly Interacting Fermions},'' \href{http://dx.doi.org/10.1103/PhysRev.111.321}{{\em Phys. Rev.} {\bfseries 111} (1958) 321--328}.

\bibitem{Landau:1960iwx}
L.~D. Landau, {\em {Fundamental Problems}}, \href{http://dx.doi.org/10.1016/b978-0-08-010586-4.50105-x}{pp.~800--802}.
\newblock 1960.

\bibitem{Chew:1962mpd}
G.~F. Chew, ``{S-Matrix Theory of Strong Interactions without Elementary Particles},'' \href{http://dx.doi.org/10.1103/RevModPhys.34.394}{{\em Rev. Mod. Phys.} {\bfseries 34} no.~3, (1962) 394--401}.

\bibitem{Chew:1957zz}
G.~F. Chew, M.~L. Goldberger, F.~E. Low, and Y.~Nambu, ``{Application of Dispersion Relations to Low-Energy Meson-Nucleon Scattering},'' \href{http://dx.doi.org/10.1103/PhysRev.106.1337}{{\em Phys. Rev.} {\bfseries 106} (1957) 1337--1344}.

\bibitem{Mandelstam:1958xc}
S.~Mandelstam, ``{Determination of the pion - nucleon scattering amplitude from dispersion relations and unitarity. General theory},'' \href{http://dx.doi.org/10.1103/PhysRev.112.1344}{{\em Phys. Rev.} {\bfseries 112} (1958) 1344--1360}.

\bibitem{Veneziano:1968yb}
G.~Veneziano, ``{Construction of a crossing - symmetric, Regge behaved amplitude for linearly rising trajectories},'' \href{http://dx.doi.org/10.1007/BF02824451}{{\em Nuovo Cim. A} {\bfseries 57} (1968) 190--197}.

\bibitem{Nambu:1969se}
Y.~NAMBU, {\em Quark model and the factorization of the Veneziano amplitude}, \href{http://dx.doi.org/10.1142/9789812795823_0024}{pp.~258--267}.
\newblock 1969.
\newblock \url{https://www.worldscientific.com/doi/abs/10.1142/9789812795823_0024}.

\bibitem{Nielsen:1970bc}
H.~B. Nielsen and P.~Olesen, ``{A Parton view on dual amplitudes},''
\href{http://dx.doi.org/10.1016/0370-2693(70)90474-0}{{\em Phys. Lett.} {\bfseries 32B} (1970) 203--206}.

\bibitem{Susskind:1970qz}
L.~Susskind, ``{Structure of hadrons implied by duality},'' \href{http://dx.doi.org/10.1103/PhysRevD.1.1182}{{\em Phys. Rev. D} {\bfseries 1} (1970) 1182--1186}.

\bibitem{Ramond:1971gb}
P.~Ramond, ``{Dual Theory for Free Fermions},'' \href{http://dx.doi.org/10.1103/PhysRevD.3.2415}{{\em Phys. Rev. D} {\bfseries 3} (1971) 2415--2418}.

\bibitem{Neveu:1971fz}
A.~Neveu and J.~H. Schwarz, ``{Tachyon-free dual model with a positive-intercept trajectory},'' \href{http://dx.doi.org/10.1016/0370-2693(71)90669-1}{{\em Phys. Lett. B} {\bfseries 34} (1971) 517--518}.

\bibitem{Gell-Mann:1964hhf}
M.~Gell-Mann, ``{The Symmetry group of vector and axial vector currents},'' \href{http://dx.doi.org/10.1103/PhysicsPhysiqueFizika.1.63}{{\em Physics Physique Fizika} {\bfseries 1} (1964) 63--75}.

\bibitem{Lee:1969fy}
T.~D. Lee and G.~C. Wick, ``{Negative Metric and the Unitarity of the S Matrix},'' \href{http://dx.doi.org/10.1016/0550-3213(69)90098-4}{{\em Nucl. Phys. B} {\bfseries 9} (1969) 209--243}.

\bibitem{Lee:1970iw}
T.~D. Lee and G.~C. Wick, ``{Finite Theory of Quantum Electrodynamics},'' \href{http://dx.doi.org/10.1103/PhysRevD.2.1033}{{\em Phys. Rev. D} {\bfseries 2} (1970) 1033--1048}.

\bibitem{Efimov:1971hgl}
G.~V. Efimov, ``{On the construction of nonlocal quantum electrodynamics},'' \href{http://dx.doi.org/10.1016/0003-4916(72)90127-3}{{\em Annals Phys.} {\bfseries 71} no.~2, (1972) 466--485}.

\bibitem{Bell:1969ts}
J.~S. Bell and R.~Jackiw, ``{A PCAC puzzle: $\pi^0 \to \gamma \gamma$ in the $\sigma$ model},'' \href{http://dx.doi.org/10.1007/BF02823296}{{\em Nuovo Cim. A} {\bfseries 60} (1969) 47--61}.

\bibitem{Adler:1969gk}
S.~L. Adler, ``{Axial vector vertex in spinor electrodynamics},'' \href{http://dx.doi.org/10.1103/PhysRev.177.2426}{{\em Phys. Rev.} {\bfseries 177} (1969) 2426--2438}.

\bibitem{Adler:1969er}
S.~L. Adler and W.~A. Bardeen, ``{Absence of higher order corrections in the anomalous axial vector divergence equation},'' \href{http://dx.doi.org/10.1103/PhysRev.182.1517}{{\em Phys. Rev.} {\bfseries 182} (1969) 1517--1536}.

\bibitem{Veltman:1968ki}
M.~J.~G. Veltman, ``{Perturbation theory of massive Yang-Mills fields},'' \href{http://dx.doi.org/10.1016/0550-3213(68)90197-1}{{\em Nucl. Phys. B} {\bfseries 7} (1968) 637--650}.

\bibitem{Reiff:1969pq}
J.~Reiff and M.~J.~G. Veltman, ``{Massive yang-mills fields},'' \href{http://dx.doi.org/10.1016/0550-3213(69)90190-4}{{\em Nucl. Phys. B} {\bfseries 13} (1969) 545--564}.

\bibitem{Bouchiat:1972iq}
C.~Bouchiat, J.~Iliopoulos, and P.~Meyer, ``{An Anomaly Free Version of Weinberg's Model},'' \href{http://dx.doi.org/10.1016/0370-2693(72)90532-1}{{\em Phys. Lett. B} {\bfseries 38} (1972) 519--523}.

\bibitem{Gross:1972pv}
D.~J. Gross and R.~Jackiw, ``{Effect of anomalies on quasirenormalizable theories},'' \href{http://dx.doi.org/10.1103/PhysRevD.6.477}{{\em Phys. Rev. D} {\bfseries 6} (1972) 477--493}.

\bibitem{Georgi:1972bb}
H.~Georgi and S.~L. Glashow, ``{Gauge theories without anomalies},'' \href{http://dx.doi.org/10.1103/PhysRevD.6.429}{{\em Phys. Rev. D} {\bfseries 6} (1972) 429}.

\bibitem{Parisi:1973ma}
G.~Parisi, ``{Deep inelastic scattering in a field theory with computable large-momenta behaviour},'' \href{http://dx.doi.org/10.1007/BF02728276}{{\em Lett. Nuovo Cim.} {\bfseries 7S2} (1973) 84--88}.

\bibitem{Callan:1973pu}
C.~G. Callan, Jr. and D.~J. Gross, ``{Bjorken scaling in quantum field theory},'' \href{http://dx.doi.org/10.1103/PhysRevD.8.4383}{{\em Phys. Rev. D} {\bfseries 8} (1973) 4383--4394}.

\bibitem{Vanyashin:1965ple}
V.~S. Vanyashin and M.~V. Terentev, ``{The Vacuum Polarization of a Charged Vector Field},'' {\em Zh. Eksp. Teor. Fiz.} {\bfseries 48} no.~2, (1965) 565--573.

\bibitem{Khriplovich:1969aa}
I.~B. Khriplovich, ``{Green's functions in theories with non-abelian gauge group.},'' {\em Sov. J. Nucl. Phys.} {\bfseries 10} (1969) 235--242.

\bibitem{Veltman:2000xp}
M.~J.~G. Veltman, ``{Nobel lecture: From weak interactions to gravitation},'' \href{http://dx.doi.org/10.1103/RevModPhys.72.341}{{\em Rev. Mod. Phys.} {\bfseries 72} (2000) 341--349}.

\bibitem{shifman2024historicalcuriosityasymptoticfreedom}
M.~Shifman, ``Historical curiosity: how asymptotic freedom of the yang-mills theory could have been discovered three times before gross, wilczek, and politzer, but was not,'' 2024.
\newblock \url{https://arxiv.org/abs/2203.12030}.

\bibitem{Hasert:1973cr}
F.~J. Hasert {\em et~al.}, ``{Search for Elastic $\nu_\mu$ Electron Scattering},'' \href{http://dx.doi.org/10.1016/0370-2693(73)90494-2}{{\em Phys. Lett. B} {\bfseries 46} (1973) 121--124}.

\bibitem{Gupta:1952zz}
S.~N. Gupta, ``{Quantization of Einstein's gravitational field: general treatment},'' \href{http://dx.doi.org/10.1088/0370-1298/65/8/304}{{\em Proc. Phys. Soc. A} {\bfseries 65} (1952) 608--619}.

\bibitem{Utiyama:1962sn}
R.~Utiyama and B.~S. DeWitt, ``{Renormalization of a classical gravitational field interacting with quantized matter fields},'' \href{http://dx.doi.org/10.1063/1.1724264}{{\em J. Math. Phys.} {\bfseries 3} (1962) 608--618}.

\bibitem{Deser:1974zz}
S.~Deser, ``{Quantum Gravity},'' in {\em {17th Int. Conf. High-Energy Physics}}, pp.~264--266.
\newblock 1974.
\newblock \url{https://inspirehep.net/files/72d8bd8268615d66e074f427ace912bb}.

\bibitem{Weinberg:1974tw}
S.~Weinberg, ``{Problems in Gauge Field Theories},'' in {\em {17th Int. Conf. High-Energy Physics}}, pp.~59--65.
\newblock 1974.
\newblock \url{https://inspirehep.net/files/deb09bc60393a3d4066a228c3b0311ef}.

\bibitem{Adler:1982ri}
S.~L. Adler, ``{Einstein Gravity as a Symmetry-Breaking Effect in Quantum Field Theory},'' \href{http://dx.doi.org/10.1103/RevModPhys.54.729}{{\em Rev. Mod. Phys.} {\bfseries 54} (1982) 729}. [Erratum: Rev.Mod.Phys. 55, 837 (1983)].

\bibitem{Strominger:1982wp}
A.~Strominger, ``{Is there a quantum theory of gravity?},'' {\em QUANTUM THEORY OF GRAVITY. Essay in honor of the 60th birthday of Bryce S. DeWitt} (12, 1982) . \url{https://drive.google.com/file/d/1Et6W1EE9aRh6aK45kgEe_wvDLZaGLv9x/view?usp=sharing}.

\bibitem{Boulware:1983yj}
D.~G. Boulware, ``{Quantization of higher derivative theories of gravity},'' {\em QUANTUM THEORY OF GRAVITY. Essay in honor of the 60th birthday of Bryce S. DeWitt} (1, 1983) . \url{https://drive.google.com/file/d/1Twen56RhzXzN1QqD_1EDnj1edG0clffg/view?usp=sharing}.

\bibitem{Boulware:1983vw}
D.~G. Boulware and D.~J. Gross, ``{LEE-WICK INDEFINITE METRIC QUANTIZATION: A FUNCTIONAL INTEGRAL APPROACH},'' \href{http://dx.doi.org/10.1016/0550-3213(84)90167-6}{{\em Nucl. Phys. B} {\bfseries 233} (1984) 1--23}.

\bibitem{Hawking:1985gh}
S.~W. Hawking, ``{WHO'S AFRAID OF (HIGHER DERIVATIVE) GHOSTS?},'' {\em Print-86-0124 (CAMBRIDGE)} (9, 1985) . \url{https://drive.google.com/file/d/17aMZOUFPX1YKhDrKKDp5hxGrtLPdqW-X/view?usp=sharing}.

\bibitem{Antoniadis:1986tu}
I.~Antoniadis and E.~T. Tomboulis, ``{Gauge Invariance and Unitarity in Higher Derivative Quantum Gravity},'' \href{http://dx.doi.org/10.1103/PhysRevD.33.2756}{{\em Phys. Rev. D} {\bfseries 33} (1986) 2756}.

\bibitem{Johnston:1987ue}
D.~A. Johnston, ``{Sedentary Ghost Poles in Higher Derivative Gravity},'' \href{http://dx.doi.org/10.1016/0550-3213(88)90555-X}{{\em Nucl. Phys. B} {\bfseries 297} (1988) 721--732}.

\bibitem{Scherk:1974ca}
J.~Scherk and J.~H. Schwarz, ``{Dual Models for Nonhadrons},'' \href{http://dx.doi.org/10.1016/0550-3213(74)90010-8}{{\em Nucl. Phys. B} {\bfseries 81} (1974) 118--144}.

\bibitem{Yoneya:1974jg}
T.~Yoneya, ``{Connection of Dual Models to Electrodynamics and Gravidynamics},'' \href{http://dx.doi.org/10.1143/PTP.51.1907}{{\em Prog. Theor. Phys.} {\bfseries 51} (1974) 1907--1920}.

\bibitem{Green:1981yb}
M.~B. Green and J.~H. Schwarz, ``{Supersymmetrical String Theories},'' \href{http://dx.doi.org/10.1016/0370-2693(82)91110-8}{{\em Phys. Lett. B} {\bfseries 109} (1982) 444--448}.

\bibitem{Green:1984sg}
M.~B. Green and J.~H. Schwarz, ``{Anomaly Cancellation in Supersymmetric D=10 Gauge Theory and Superstring Theory},'' \href{http://dx.doi.org/10.1016/0370-2693(84)91565-X}{{\em Phys. Lett. B} {\bfseries 149} (1984) 117--122}.

\bibitem{Gross:1984dd}
D.~J. Gross, J.~A. Harvey, E.~J. Martinec, and R.~Rohm, ``{The Heterotic String},'' \href{http://dx.doi.org/10.1103/PhysRevLett.54.502}{{\em Phys. Rev. Lett.} {\bfseries 54} (1985) 502--505}.

\bibitem{Candelas:1985en}
P.~Candelas, G.~T. Horowitz, A.~Strominger, and E.~Witten, ``{Vacuum configurations for superstrings},'' \href{http://dx.doi.org/10.1016/0550-3213(85)90602-9}{{\em Nucl. Phys. B} {\bfseries 258} (1985) 46--74}.

\bibitem{Freedman:1976xh}
D.~Z. Freedman, P.~van Nieuwenhuizen, and S.~Ferrara, ``{Progress Toward a Theory of Supergravity},'' \href{http://dx.doi.org/10.1103/PhysRevD.13.3214}{{\em Phys. Rev. D} {\bfseries 13} (1976) 3214--3218}.

\bibitem{Weinberg:1976xy}
S.~Weinberg, \href{http://dx.doi.org/10.1007/978-1-4684-0931-4_1}{``{Critical Phenomena for Field Theorists},''} in {\em {14th International School of Subnuclear Physics: Understanding the Fundamental Constitutents of Matter}}.
\newblock 8, 1976.

\bibitem{Weinberg:1980gg}
S.~Weinberg, {\em ULTRAVIOLET DIVERGENCES IN QUANTUM THEORIES OF GRAVITATION}, pp.~790--831.
\newblock 1980.
\newblock \url{https://drive.google.com/file/d/1tXdTgiA9lGx_cGaR7txXzjwJX1s41Rjx/view?usp=sharing}.

\bibitem{Donoghue:2019clr}
J.~F. Donoghue, ``{A Critique of the Asymptotic Safety Program},'' \href{http://dx.doi.org/10.3389/fphy.2020.00056}{{\em Front. in Phys.} {\bfseries 8} (2020) 56}, \href{http://arxiv.org/abs/1911.02967}{{\ttfamily arXiv:1911.02967 [hep-th]}}.

\bibitem{Ashtekar:1986yd}
A.~Ashtekar, ``{New Variables for Classical and Quantum Gravity},''
\href{http://dx.doi.org/10.1103/PhysRevLett.57.2244}{{\em Phys. Rev. Lett.} {\bfseries 57} (1986) 2244--2247}.

\bibitem{Ashtekar:1987gu}
A.~Ashtekar, ``{New Hamiltonian Formulation of General Relativity},''
\href{http://dx.doi.org/10.1103/PhysRevD.36.1587}{{\em Phys. Rev.} {\bfseries D36} (1987) 1587--1602}.

\bibitem{Rovelli:1989za}
C.~Rovelli and L.~Smolin, ``{Loop Space Representation of Quantum General Relativity},''
\href{http://dx.doi.org/10.1016/0550-3213(90)90019-A}{{\em Nucl. Phys.} {\bfseries B331} (1990) 80--152}.

\bibitem{Bombelli:1987aa}
L.~Bombelli, J.~Lee, D.~Meyer, and R.~Sorkin, ``{Space-Time as a Causal Set},'' \href{http://dx.doi.org/10.1103/PhysRevLett.59.521}{{\em Phys. Rev. Lett.} {\bfseries 59} (1987) 521--524}.

\bibitem{Krasnikov:1987yj}
N.~V. Krasnikov, ``{NONLOCAL GAUGE THEORIES},'' \href{http://dx.doi.org/10.1007/BF01017588}{{\em Theor. Math. Phys.} {\bfseries 73} (1987) 1184--1190}.
[Teor. Mat. Fiz.73,235(1987)].

\bibitem{Kuzmin:1989sp}
{\relax Yu}.~V. Kuzmin, ``{THE CONVERGENT NONLOCAL GRAVITATION. (IN RUSSIAN)},'' {\em Sov. J. Nucl. Phys.} {\bfseries 50} (1989) 1011--1014.
[Yad. Fiz.50,1630(1989)].

\bibitem{WMAP:2012nax}
{\bfseries WMAP} Collaboration, G.~Hinshaw {\em et~al.}, ``{Nine-Year Wilkinson Microwave Anisotropy Probe (WMAP) Observations: Cosmological Parameter Results},'' \href{http://dx.doi.org/10.1088/0067-0049/208/2/19}{{\em Astrophys. J. Suppl.} {\bfseries 208} (2013) 19}, \href{http://arxiv.org/abs/1212.5226}{{\ttfamily arXiv:1212.5226 [astro-ph.CO]}}.

\bibitem{Planck:2013jfk}
{\bfseries Planck} Collaboration, P.~A.~R. Ade {\em et~al.}, ``{Planck 2013 results. XXII. Constraints on inflation},'' \href{http://dx.doi.org/10.1051/0004-6361/201321569}{{\em Astron. Astrophys.} {\bfseries 571} (2014) A22}, \href{http://arxiv.org/abs/1303.5082}{{\ttfamily arXiv:1303.5082 [astro-ph.CO]}}.

\bibitem{Loll:1998aj}
R.~Loll, ``{Discrete approaches to quantum gravity in four-dimensions},'' \href{http://dx.doi.org/10.12942/lrr-1998-13}{{\em Living Rev. Rel.} {\bfseries 1} (1998) 13}, \href{http://arxiv.org/abs/gr-qc/9805049}{{\ttfamily arXiv:gr-qc/9805049}}.

\bibitem{Kruczenski:2022lot}
M.~Kruczenski, J.~Penedones, and B.~C. van Rees, ``{Snowmass White Paper: S-matrix Bootstrap},'' \href{http://arxiv.org/abs/2203.02421}{{\ttfamily arXiv:2203.02421 [hep-th]}}.

\bibitem{Butterfield:2014rxa}
J.~Butterfield and N.~Bouatta, ``{Renormalization for Philosophers},'' \href{http://arxiv.org/abs/1406.4532}{{\ttfamily arXiv:1406.4532 [physics.hist-ph]}}.

\bibitem{Butterfield:2014oja}
J.~Butterfield, ``{Reduction, Emergence and Renormalization},'' {\em J. Philos.} {\bfseries 111} (2014) 5--49, \href{http://arxiv.org/abs/1406.4354}{{\ttfamily arXiv:1406.4354 [physics.hist-ph]}}.

\end{thebibliography}\endgroup


\end{document}